\documentclass [10pt]{article}
\usepackage{amssymb}
\usepackage{amsmath}
\usepackage{graphicx}
 
\usepackage{mathptmx} 

\newcommand{\bi}{\begin{itemize} \vspace*{-0.1in}}
\newcommand{\be}{\begin{equation}}
\newcommand{\ee}{\end{equation}} 
\newcommand{\ba}{\begin{equation} \begin{aligned}}
\newcommand{\ea}{\end{aligned} \end{equation}}
\newcommand{\bracket}[1]{\left\langle{#1}\right\rangle}

\begin{document}

\title{Effects of rapid prey evolution on predator-prey cycles} 
\author{Laura E. Jones and Stephen P. Ellner\\
       Ecology and Evolutionary Biology, \\
         \textsc{Cornell University} \\ 
             Ithaca NY 14853 \\  
             Tel.: +01-607-254-4268              
}

\maketitle
\begin{abstract}
\small
We study the qualitative properties of population cycles in a predator-prey 
system where genetic variability allows contemporary rapid evolution of the prey. Previous numerical
studies have found that prey evolution in response to changing predation risk 
can have major quantitative and qualitative effects on predator-prey cycles, including:
(i) large increases in cycle period, (ii) changes in phase relations (so that predator and prey
are cycling exactly out of phase, rather than the classical quarter-period phase lag),
and (iii) ``cryptic'' cycles in which total prey density remains nearly constant while predator
density and prey traits cycle. Here we focus on a chemostat model motivated by
our experimental system \cite{Fussmann2000,Yoshida2003} with algae (prey) and rotifers (predators), 
in which the prey exhibit rapid evolution in their level of defense against predation.  We show that the
effects of rapid prey evolution are robust and general, and furthermore that they occur in a specific but biologically relevant region of parameter space: when traits that greatly reduce predation risk are relatively cheap (in terms of reductions in other fitness components), when there is coexistence between the two prey types and the predator, and  when the interaction between predators and undefended prey alone would produce cycles.  Because defense has been shown to be inexpensive,
even cost-free, in a number of systems 
\cite{Andersson1999,Gagneux2006,Yoshida2004}, our discoveries may well be reproduced in other model systems, and in nature.
Finally, some of our key results are extended to a general model in which functional forms for the predation rate and prey birth rate are not specified.
\end{abstract}
\normalsize
\pagebreak
\section{Introduction}
The potential for rapid evolutionary change is well documented: in some natural and experimental settings, trait evolution and population dynamics are observed to occur on similar time scales \cite{Hairston2005,Hendry1999,Palumbi2001,Thompson1998}.  Though laboratory demonstrations of the interactions between trait and population dynamics still remain rare, there are a few direct examples \cite{Conover2002,Meyer2006} and also examples where these interactions are inferred via modeling (e.g., \cite{Bohannan1999,Yoshida2003,Yoshida2006}). Furthermore,  documented cases where the effects of rapid evolution are observed in a changing natural environment are increasing (e.g., 
\cite{Antonovics1971,Coltman2003,Grant2002,Hairston1986,Heath2003,Olsen2004,Reznick1997}, 
and for reviews on the topic, 
\cite{Ashley2003,Hairston2005,SaccheriHanski2006,Zimmer2003}).
Among the fishes, rapid evolution has been observed or inferred
in natural populations of atlantic cod ({\it Gadus morhua}), which  evolved towards reproductive maturity at earlier ages and smaller sizes under heavy size-dependent selection (fishing) \cite{Olsen2004}.  Within four hatchery generations, Chinook salmon ({\it Oncorhynchus tshawytscha}) raised in hatcheries to supplement natural populations evolve life-history traits (smaller eggs) that reduce odds of survival in the wild \cite{Heath2003}. 

Where prey are under strong selection to avoid predation (because the risk of getting
eaten is very strong natural selection),
prey genetic diversity yields the capacity for rapid evolution of resistance to predation, akin to the rapid evolution of microbial
pathogens in response to antibiotics. However,  prey defensive strategies, be they
physiological, behavioral, or developmental,
may themselves exact demographic costs from the prey equivalent to or greater than
that of direct consumption by predators \cite{Preisser2005}. 
The energetic costs of defense traits in ``trait mediated interactions",
for example the development of armored spines in {\it Daphnia} when exposed to chemicals
from fish, reduce lifetime fitness by routing energy to defense which would
otherwise go into progeny \cite{Barry1994}.    Thus, by 
focusing exclusively on changes in population {\it numbers}
without considering changes in the {\em properties} (and associated costs) of individuals comprising the populations, conventional models of 
population and community dynamics may give us, at best, only half the story in a system where prey adapt to predation risk. The capacity for rapid evolution may expand the conditions permitting coexistence of predator and prey. For example, within the ecosystem of the human body, rapid prey evolution may allow viruses (prey) to escape detection by immune cells (predators) by altering a few surface proteins, or allow a pathogenic strain of bacteria to evolve so that it is no longer sensitive to a widely-used antibiotic.  Heterogeneity in prey edibility may also shift prey population regulation from top-down to bottom-up by providing a haven from predation in the form of the less edible prey.  

\begin{figure}[t]
\centerline{\includegraphics[width=5.0in]{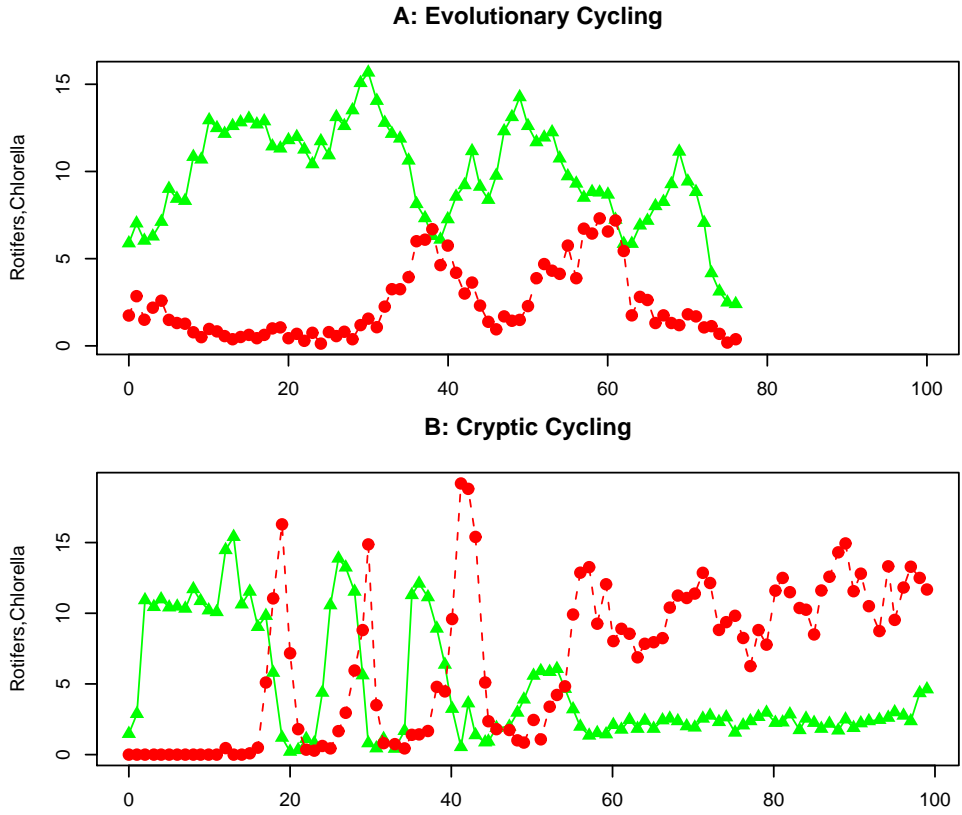}}
\caption{\small{A, ``Evolutionary'' cycling.  {\it Chlorella} are shown in solid line, and
the rotifer predator is shown dashed.  Phase relations are nearly out-of-phase, unlike the 
quarter-phase shift seen in classic predator-prey cycles. B,  Cryptic Cycling.  Initially the system exhibits
classic predator-prey cycles, which would be expected when a single (edible) prey type is dominant.  At about day 50
the system switches to cryptic cycles, which would be expected if a highly defended (inedible) type with low cost for
defense arose by mutation. A switch from classic to cryptic cycles when a defended type arises by mutation has been
documented in bacteria-phage chemostats \cite{Bohannan1999}.}
}
\label{ExampleCycles}
\end{figure}

Our experimental system is a predator-prey microcosm with rotifers, 
\textit{Brachionus calyciflorus}, and their algal prey, \textit{Chlorella vulgaris}, cultured together in nitrogen-limited, continuous flow-through chemostats.     
In prior studies, we have shown that coexistence of edible and inedible prey types
(genetic variation in the algal prey) allows the prey to evolve in response to temporally variable selection due to predation pressure from the rotifer predator,  and nutrient limitation at high prey densities. 
While our earlier studies did not specifically track changes at the genotypic level, recent work on our system explicitly identifies two competing algal strains, and tracks changes in their densities as the algal population evolves under predation pressure \cite{Meyer2006}.  Evolution in the prey can lead to ``evolutionary'' cycling 
\cite{Schertzer2002,Yoshida2003}, where the predator and prey
exhibit extended, out-of-phase population cycles (Figure \ref{ExampleCycles}A), or in some circumstances, 
the odd phenomenon of ``cryptic cycles", where the predator alone exhibits regular
population cycles but the prey appear to remain in steady state (Figure \ref{ExampleCycles}B).  
In cryptic cycling, densities of edible and inedible prey cycle out of phase with each other, driven by changes in predator abundance, in such a way that total prey density 
remains nearly constant \cite{Yoshida2006}.   These dynamics are not unique to the organisms
in this system: we have observed evolutionary cycles in a chemostat system
comprised of rotifers cultured with the flagellated algae {\it Chlamydomonus}, 
and cryptic dynamics have been observed in
bacteria-phage microcosms \cite{Bohannan1999,Yoshida2006}.    
We are motivated here by just these sorts of perplexing experimental results 
from our system, and by the close match between our experimental data and model simulations. 

Understanding the potential effects of rapid evolution on the dynamics of natural ecosystems is critical to predicting how populations will adapt to a changing environment. 
Populations in the wild today face unprecedented stress from habitat loss or degradation, harvesting pressure, species introductions and climate change.  
In addition, otherwise well-intentioned attempts at conservation or management often
fail to take into account the potential for rapid Darwinian responses
to intervention \cite{KinnisonHairston2006}.
Thus before conclusions based on laboratory systems or manipulated natural systems are applied to the natural world, we must ask if the conclusions are likely to be robust: are they limited to the special conditions present in the experimental systems, or should we expect to see them in a broad range of conditions in nature? The present contribution is an attempt, using theory, to answer the questions: how general is the phenomenon of evolutionary cycling in predator-prey systems, under what circumstances might these dynamics be observed, and what are the implications of this type of phenomenon for natural systems?

\section{The model}

Our model is based on an experimental predator-prey microcosm with rotifers, 
\textit{Brachionus calyciflorus}, and their algal prey, \textit{Chlorella vulgaris}, cultured together in a nitrogen-limited, continuous flow-through chemostat system. This system was first described by Fussmann et al. \cite{Fussmann2000}, further characterized by Schertzer et al. \cite{Schertzer2002} and  Yoshida et al. \cite{Yoshida2003,Yoshida2004}, and equilibrium properties studied by Jones and Ellner \cite{JonesEllner2004}. 
\textit{Brachionus} in the wild are facultatively sexual, but because sexually produced eggs wash out of the chemostat before offspring hatch, our rotifer cultures have evolved to be entirely parthenogenic \cite{Fussmann2003}. The algae also reproduce asexually \cite{PickettHeaps1975}, so evolutionary change in the prey occurs as a result
of changes in the relative frequency of different algal clones.

We use a system of ordinary differential equations to describe the population and prey evolutionary dynamics in the experimental microcosms \cite{Yoshida2003,JonesEllner2004}. 
Genetic variability and thus the possibility of evolution in the prey is introduced by explicitly representing the prey 
population as a finite set of asexually reproducing clones. Each clone is 
characterized by its palatability $p$, which represents the conditional 
probability that an algal cell is digested rather than being ejected alive, once it has been ingested
by a predator \cite{Meyer2006}. 

The model consists of three equations for the limiting nutrient and
rotifers, plus $q$ equations for $q$ prey clones. In the following equations,
$N$ is nitrogen ($\mu$ mol per liter), $C_i$ represents concentration  of
the $i^{th}$ algal clone ($10^9$ cells per liter), where 
$i = 1,2, ....,q$. Here we limit the number of clones to two, for reasons
discussed below. $R$ and $B$ are the fertile and total population densities, respectively, 
for the predator \textit{Brachionus} (individuals per liter). Fertile rotifers 
senesce and stop breeding at rate $\lambda$; all rotifers are subject to fixed mortality $m$.
The parameters $\chi_c, \chi_b$ are conversions between consumption and recruitment
rates (additional model parameters are defined in Table 1). 

\be
\begin{aligned}
\frac{dN}{dt} &  =  \delta(N_I -  N)  - \sum_{i=1}^{q}F_{C,i}(N)C_i  \\
\frac{dC_i}{dt} & =  \chi_c F_{C,i}(N)C_i  - p_i F_{b}(C_i)B - \delta C_i \\
\frac{dR}{dt} & =  \chi_b\sum_{i=1}^{2}p_i F_{b}(C_i)R - (\delta + m + \lambda)R\\
\frac{dB}{dt} & =  \chi_b\sum_{i=1}^{2}p_i F_{b}(C_i)R- (\delta + m)B 
\end{aligned}
\label{eq1} 
\ee

\noindent
where $$ F_{c,i}(N)= \rho_c N/(K_c(p_i) + N) $$

\noindent 
and 
\be
\label{eq3}
 F_{b}(C_i)=G C_i/(K_b + \sum\limits_{i=1}^{2}C_i p_i)
\ee

\noindent
are functional response equations describing algal and rotifer consumption rates, respectively, 
and where $\rho_c = \omega_c\beta_c/\epsilon_c$.
Equation \eqref{eq3} is derived from the predator's clearance rate \textbf{G} 
(the volume of water per unit time that an individual filters to obtain food). 
We assume that clearance rate is a function of the total prey food value:
\be
\textbf{G}=\frac{G}{K_b + \sum\limits_{i=1}^{2}C_i p_i}.
\label{Gmodel}
\ee
That is, lower prey palatability results in the predators increasing their clearance rate, 
exactly as if prey were less nutritious. We also considered a model in which clearance rate depends 
only on the total prey density, but it could not be fitted as well to our experimental 
data on population cycles. Elsewhere \cite{Yoshida2003,Meyer2006} we have used a more
complicated expression for \textbf{G} in order to fit experimental data more accurately, but using
\eqref{Gmodel} does not change the model's qualitative behavior. 

The cost for defense against predation is a reduced ability to compete for scarce nutrients 
\cite{Yoshida2004,JonesEllner2004,Meyer2006}. We model this by specifying a tradeoff curve 
\be
\label{tradeoff}
K_c(p) = K_c + \alpha_2(1-p)^{\alpha_1}. 
\ee
Here $K_c>0$ is the minimum value of the half-saturation constant, $\alpha_1>0$ determines whether 
the tradeoff curve is concave up versus down, and $\alpha_2 > 0$ is the cost for becoming
completely inedible ($p=0$). 

\begin{table}[t]
\caption{Parameter estimates for the \textit{Chlorella-Brachionus} microcosm system.
Set: adjustable parameters set by the experimenter.
TY: Unpublished experimental data (Yoshida \textit{et al.}, in prep.) 
Fitted: Estimated by numerically optimizing the goodness-of-fit between model output
and data on total prey and predator population dynamics from two experiments 
(originally reported by Fussmann \cite{Fussmann2000}) in which regular cycles occurred.}
\centering
\label{Table 1.}       
\begin{tabular}{llll}
\hline\noalign{\smallskip}
Parameter & Description & value & Reference \\[3pt]
\hline
$N_I$ & Limiting nutrient conc. (suplied medium) & $80 \mu$ mol N$/l$ & Set \\
$\delta$ & Chemostat dilution rate & variable ($d$) &  Set  \\
$V$ & Chemostat volume & $0.33 l$ & Set  \\
$\chi_c$ & Algal conversion efficiency ($10^9$ cells/$\mu$mol N) & 0.05 & \cite{Fussmann2000}   \\
$\chi_b$ & Rotifer conversion efficiency & $\approx 54000$ rotifers/$10^9$ algal cells & Fitted \\
$m$ & Rotifer mortality & $0.055 /d$ &  \cite{Fussmann2000} \\
$\lambda$ & Rotifer senescence rate & $0.4 /d$ & \cite{Fussmann2000} \\
$K_c$ & Minimum algal half-saturation & $4.3$ $\mu$ mol N $/l$  & \cite{Fussmann2000} \\
$K_b$ & Rotifer half-saturation & $0.835 \times 10^9$ algal cells $/l$ &  TY \\
$\beta_c$ & Maximum algal recruitment rate & $3.3 /d$ & TY  \\
$\omega_c$ & N content in $10^9$ algal cells & $20.0$ $\mu$ mol & \cite{Fussmann2000}  \\
$\epsilon_c$ & Algal assimilation efficiency & $1$ & \cite{Fussmann2000}  \\
$G$ & Rotifer maximum consumption rate& $5.0\times10^{-5}$ $l/d$  & TY \\
$\alpha_1$ & Shape parameter in algal tradeoff  & variable, $\alpha_1 > 0$ & Fitted \\
$\alpha_2$ & Scale parameter in algal tradeoff  & variable, $\alpha_2 > 0$ & Fitted \\
\noalign{\smallskip}\hline
\end{tabular}
\end{table}

\section{Characteristics of the model under simulation}
A system of $q > 2$ prey types invariably collapses to 
at most two types in the presence of a predator: either a single clone
that outcompetes all others, or a pair of very different clones 
(one very well defended and the other highly competitive) that together 
drive all intermediate prey types to extinction \cite{Yoshida2003,JonesEllner2004}. 
Only the latter case is of interest here, because with a single prey type there is 
no prey evolution. We thus consider here a system of two extreme prey types in 
the presence of a predator.

Two system parameters can be experimentally varied: 
the dilution rate $\delta$
(fraction of the culture medium that is replaced daily) and the concentration of the limiting nutrient in the 
inflowing medium, $N_I$. Fussmann et al. \cite{Fussmann2000} showed that $\delta$ is a bifurcation
parameter: in both the real system and the model, the system goes to equilibrium at low dilution rates, 
limit cycles at intermediate dilution rates, and again to equilibrium at high dilution rates. Further increases
in $\delta$ lead to extinction of the predator. Toth and Kot \cite{TothKot2006} proved that 
the same bifurcation sequence occurs in chemostat models with an age-structured consumer feeding
on an abiotic resource (for our experimental system, this would be rotifers feeding on externally-supplied
algae that could not reproduce within the chemostat). 

The prey vulnerability parameter $p$ is also a bifurcation parameter. In the following discussion, we define 
\textit{evolutionary cycles} as both prey types coexisting and exhibiting long-period cycles (period 20--40 days), with the
predator and total prey abundance almost exactly out-of-phase with each other. \textit{Predator-prey cycles} 
are shorter (6--12 days), display the classic quarter-period phase offset between predator and prey, and involve one prey type cycling with the predator. In addition, both prey may survive and coexist with the predator at an \textit{evolutionary equilibrium}, or one prey type may be driven to extinction while the other goes to equilibrium 
with the predator.

\begin{figure}[t]
\centerline{\includegraphics[width=6in]{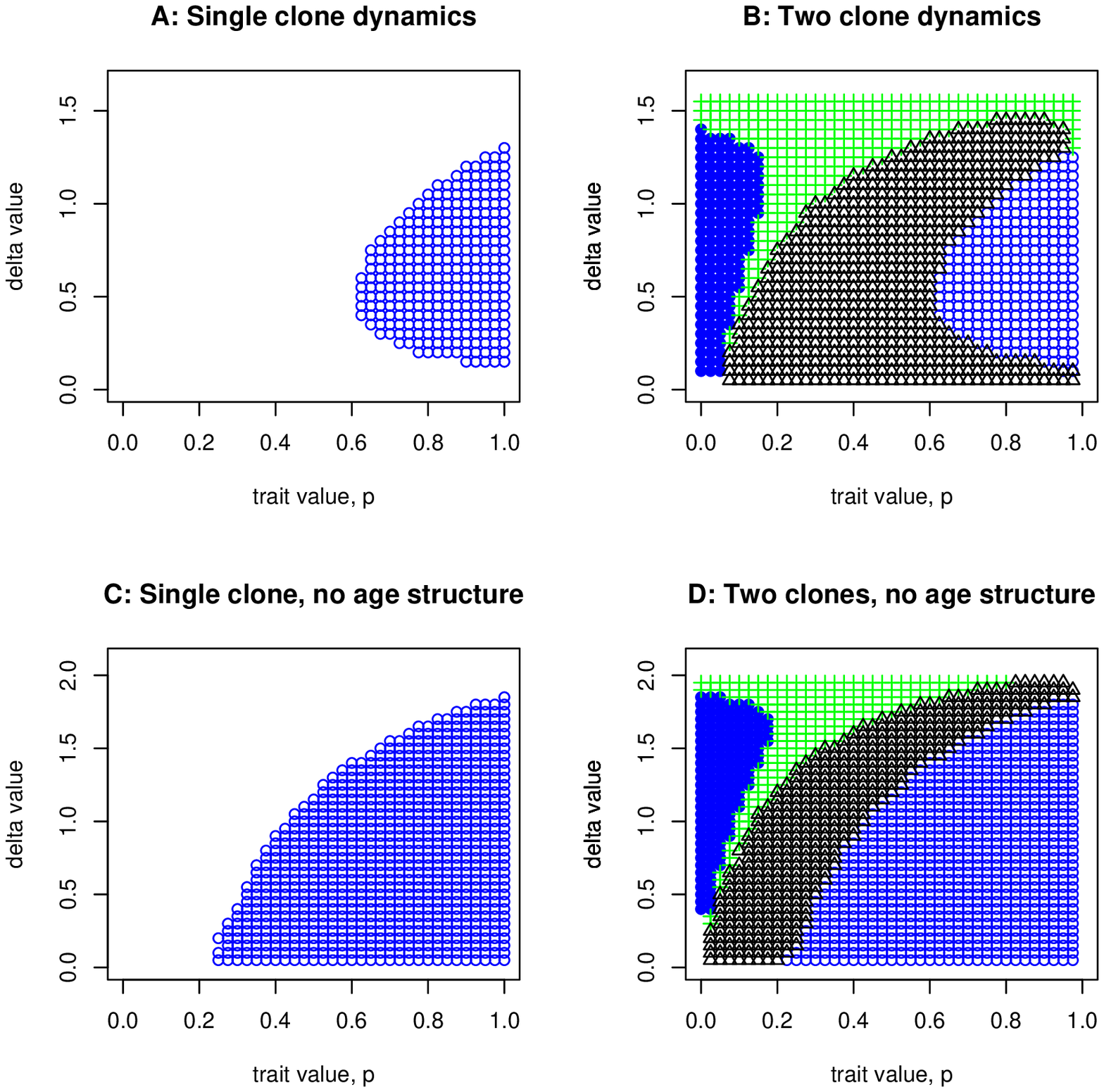}}
\caption{\small Dynamics of one- and two-prey models as a function of the palatability $p$
of the defended prey type and the dilution rate $\delta$. Panels A and B show results for the ``full''
model including predator age-structure; panels C and D show results for the ``reduced'' model without
predator age structure. A,C. Dynamics of a single prey system. Open circles show 
predator-prey cycles, and white space indicates equilibrium. B,D. Dynamics of a two-prey system. The model
is scaled so that the undefended prey type has $p=1$. Filled circles indicate that both types coexist and cycle together. Open circles show short predator-prey cycles with only the defended type ($p<1$). Cross-hatching indicates the defended and undefended 
prey coexisting at a stable equilibrium. Open triangles indicate equilibrium between predator and defended prey and white space indicates equilibrium between predator and vulnerable type.  In the model with age structure (panel B), predator extinction occurs for $\delta > 1.5 $}. 
\label{DynamicsFigs}
\end{figure}

\paragraph{Single prey model} 
Figure \ref{DynamicsFigs}A shows the dynamics of the single prey model as a function of prey 
palatability $p$ and dilution rate $\delta$. Parameters giving single-prey predator-prey cycles are indicated by open circles, and elsewhere the system goes to equilibrium. At low $p$ values (up to 0.4--0.6, depending on the predator 
conversion efficiency $\chi_B$) the system goes to equilibrium at all dilution rates. As $p$ increases 
there is a bifurcation and short, low amplitude predator-prey cycles are observed, initially for the narrow 
range of dilution rates. 
When $p$ is higher, oscillations grow in amplitude and increase very slightly in period, and cycling occurs over 
a larger range of dilution rates. The cycles always exhibit classic predator-prey phase relations. 

\paragraph{Two prey models} 
Figure \ref{DynamicsFigs}B shows dynamics of the two prey model as a function of the
dilution rate $\delta$ and the trait value $p$  of the defended prey type (the model is scaled so that the 
undefended type has $p=1$). 
Using the parameter values listed in Table 1, 
extended evolutionary cycles (closed circles) initially appear for all dilution rates
($0.2 \le \delta \le 1.3$  at $p_1$ very small ($p_1 \approx 0.01$). As $p_1$ increases,
evolutionary cycling occurs for a diminishing range of dilution rates. By $p_1 \doteq 0.2 $, cycling vanishes 
and instead the defended prey is in equilibrium (Figure \ref{DynamicsFigs}B , white space) or the two prey types are
in an evolutionary equilibrium with the predator (Figure \ref{DynamicsFigs}B , crosshatching).
As $p_1$ increases further ($0.4 < p_1 < 0.6$), 
there is another bifurcation and the system, comprised of the defended type and the
predator, begins to exhibit predator-prey cycles (Figure 
\ref{DynamicsFigs}B, open circles).
From this point on the system behaves as if it were dominated by the 
defended type (see above), until $p_1$ has increased to the point that the two
prey types are almost identical. At that point there are predator-prey cycles
with both prey types present (closed circles) but these appear to be very long 
transients rather than indefinite coexistence: one or the other prey type, depending 
on the dilution rate, is slowly driven to extinction by its competitor. 

\paragraph{Eliminating predator age structure}
Panels C and D in Figure \ref{DynamicsFigs} shows model dynamics  
{\em without} age structure in the predator. 
Age structure is removed by setting
$\lambda=0$ in \eqref{eq1}, with all other parameters unchanged.
As seen in Figure \ref{DynamicsFigs}C, the single prey model without 
age structure exhibits dynamics very similar to those in 
\ref{DynamicsFigs}B, where age structure is included.  
Predator age structure is generally stabilizing in this model because senescent rotifers are a resource sink, eating
prey without converting them to offspring. This effect is most pronounced at low values of $\delta$ because
senscent rotifers then spend more time in the chemostat before getting washed out.  
Omitting age structure is therefore destabilizing: it permits cycles with better
defended prey (lower $p$) and eliminates entirely the stability at very low $\delta$ for nearly 
all $p$ values. Similarly, simulations of the two-prey model show that eliminating predator
age structure shifts the region of ($p, \delta$) values giving evolutionary cycles to higher dilution 
rates, and eliminates the stabilization at very low $\delta$, but otherwise the bifurcation diagram is unchanged. 
Given these similarities in model behavior, we may simplify model \eqref{eq1} by eliminating predator 
age-structure without changing the properties of interest for this paper.

\section{Rescaling the model}
We now simplify the model (\ref{eq1}-\ref{eq3}) by rescaling and further reducing its order.
Based on the simulation results above we omit age structure in the predator, which is now represented by the one state
variable $B$. We also assume that predator mortality $m$ is negligibly small
relative to washout at the dilution rates $\delta$ of interest, and set $m=0$.  
The model then becomes:
\be
\begin{aligned}
\frac{dN}{dt} & =  \delta(N_I - N) - \rho_c \sum_{i=1}^2 \frac{N C_i}{K_c(p_i) + N}  \\
\frac{dC_i}{dt}& = C_i\left[\chi_c\rho_c\frac{N}{K_c(p_i)+ N}-\frac{Gp_iB}{(K_B + \sum p_i C_i)}-\delta\right]\\
\frac{dB}{dt}& =  B\left[\chi_b\frac{G\sum p_i C_i}{K_B + \sum p_i C_i} - \delta \right]. 
\end{aligned}
\label{eqs4}
\ee
We order the prey types so that $p_1$ and $p_2$ correspond to the {\em defended} and {\em vulnerable} prey types, 
respectively, $p_1 < p_2$. The cost for defense is reduced ability to compete
for scarce nutrients, so $K_c(p_1) > K_c(p_2)$. 

To rescale the model we make the following transformations:
\be
S=\frac{N}{N_I}, \quad x_i = \frac{C_i}{\chi_cN_I}, 
\quad y= \frac{B}{\chi_c\chi_b N_I}, \quad g=\frac{\chi_bG}{\delta}, 
\quad m= \frac{\chi_c\rho_c}{\delta}, \quad k_b=\frac{K_B}{\chi_c N_I}.
\quad \tau=\delta t 
\ee
The half-saturation constants for each of the two prey types are transformed as follows, 
\be
k_1=K_c(p_1)/{N_I}, \quad k_2= {K_c(p_2)}/{N_I}. 
\ee
Substituting these into \eqref{eqs4} gives:
\be
\begin{aligned}
\dot S & =  1-S - S\sum_{i=1}^2 \frac{m x_i}{k_i + S}  \\
\dot x_i & = x_i\left[\frac{m S}{k_i + S}-\frac{g p_i y}{(k_b + Q)}-1\right]\\
\dot y & =  y\left[\frac{gQ}{k_b + Q} - 1\right] 
\end{aligned}
\label{eqs5}
\ee
where 
\be
Q=p_1x_1 + p_2x_2.
\label{defineQ}
\ee
Table 2 gives values of the rescaled model parameters corresponding
to the parameter estimates in Table 1. 

\begin{table}[t]
\caption{Estimates of rescaled model parameters 
for the \textit{Chlorella-Brachionus} microcosm system.}
\centering
\label{Table 2}       
\begin{tabular}{lll}
\hline\noalign{\smallskip}
Parameter & Description & Value \\[3pt]
\hline
$m$ & Algal maximum per-capita population growth rate & 3.3/$\delta$ \\
$k_1,k_2$ & Algal half-saturation constants for nutrient uptake & 0.054 \\
$g$ & Predator maximum grazing rate & 2.55/$\delta$ \\
$k_b$ & Predator half-saturation constant for prey capture &0.21 \\
\noalign{\smallskip}\hline
\end{tabular}
\end{table}

We can reduce the dimension of the system further by letting
$\Sigma=S+x_1+x_2+y.$
From \eqref{eqs5} we have  $\dot\Sigma=1-\Sigma,$
so $\Sigma(t) \rightarrow 1 $ quickly as
$t \rightarrow \infty.$  Thus, to study the long-term dynamics of \eqref{eqs5},
we may consider the dynamics on the invariant set $\Sigma \equiv 1.$
Then $S(t)= 1- x_1(t) -x_2(t) - y(t)$, 
and \eqref{eqs5} reduces to
\be
\begin{aligned}
\dot x_i & = x_i\left[\frac{m (1 - X - y)}{k_i + (1-X-y)}-\frac{g p_i y}{(k_b + Q)}-1\right], \quad i=1,2\\
\dot y & =  y\left[\frac{gQ}{k_b + Q} - 1\right] 
\end{aligned}
\label{eqs6}
\ee
where $X=x_1 + x_2.$

\section{Analysis}
Our goals in this section are to find the conditions under which two prey 
types can coexist, to determine when coexistence is steady-state versus oscillatory, and 
to characterize the period of cycles and the phase relations during
oscillatory coexistence and during transients when one type is decreasing to extinction.  
Throughout this section we consider the reduced model \eqref{eqs6}. 
For local stability analysis it is useful to note that the model has
the form 
\be
\dot x_i = x_i r_i(x_1,x_2,x_3) \quad i=1,2,3
\label{GeneralForm}
\ee
with $x_3=y$. It follows that at any equilibrium where the $x_i$ are all positive (and
hence the $r_i$ are all 0) the Jacobian matrix $J$ has entries 
\be
J(i,j)=\tilde x_i \frac{\partial \tilde r_i}{\partial x_j}
\label{GeneralJacobian}
\ee
with the tilde indicating evaluation at the equilibrium with all $x_i$ present. It
is also useful for local stability analysis that the determinant of \eqref{GeneralJacobian}
is always negative unless $p_1=p_2$ (Appendix \ref{JacobianAppendix}). 

\subsection{Dynamics of a one-prey system}
We need first some properties of the one-prey model
\be
\begin{aligned}
\dot x & = x\left[\frac{m (1 - x - y)}{k + (1-x-y)}-\frac{g p y}{(k_b + px)}-1 \right]\\
\dot y & =  y \left[\frac{gpx}{k_b + px} - 1 \right]. 
\end{aligned}
\label{eq7}
\ee
This is a standard predator-prey chemostat model and its behavior is well-known,
so we summarize here only the results that we will need later; 
see e.g. \cite{SmithWaltman1995} for derivations and details. 

In the absence of predators, 
the steady state for this system is 
$
E_0= (1-\Lambda,0), 
$
where 
\be
\Lambda= \frac{k}{m-1}.
\label{Lambda}
\ee
$\Lambda$ is the scaled concentration of limiting nutrient at
which prey growth balances washout rate, so that $\dot x = 0.$
Similarly, steady state densities for each prey type in a predator-free two clone system are
$$
E_i= (1-\Lambda_i,0) \quad  \textrm{where}\quad \Lambda_i=\frac{k_i}{m-1}.
$$  
The steady state for the prey in the presence of the predator is
\be
\label{xc}
\bar x_c = \frac{k_b}{p(g-1)};
\ee
$\bar x_c$ is the prey density at which the predator birth and death rates are equal. 
The model \eqref{eq7} has an interior equilibrium point $E_c=(\bar x_c, \bar y_c)$ representing 
predator-prey coexistence if  
\be
\Lambda + x_c < 1 
\label{coexistence2}
\ee
\cite{SmithWaltman1995}, and otherwise the predator cannot persist. 
The system then collapses to the prey by itself 
and converges to $E_0$. Condition \eqref{coexistence2} says that
there is an interior equilibrium if the prey by themselves reach a steady state
($1-\Lambda$) that provides enough food so that the predator birth rate exceeds
the predator death rate.   

The expression for the steady state of the predator, $y_c$, is easily obtained from \eqref{eq7}: 
\be
\label{predatorSS}
\bar y_c=\bar \sigma -\bar x_c,\quad \mbox{where} \quad \bar\sigma = 
(\bar x_c + \bar y_c) = \frac{1}{2}\left[\gamma - \sqrt{(\gamma^2 - 4m\bar x_c)}\right], 
\ee
with $\gamma=k + 1 + m\bar x_c$. Similarly, the steady-state densities for the predator in a single-prey system with either prey type, $\bar y_i$, are found by substituting the steady state for the prey, $\bar x_i$, 
in place of $\bar x_c$ and the appropriate half-saturation
$k_i$ in place of $k$ in \eqref{predatorSS}.

\begin{figure}[t]
\centerline{\includegraphics[width=3.5in]{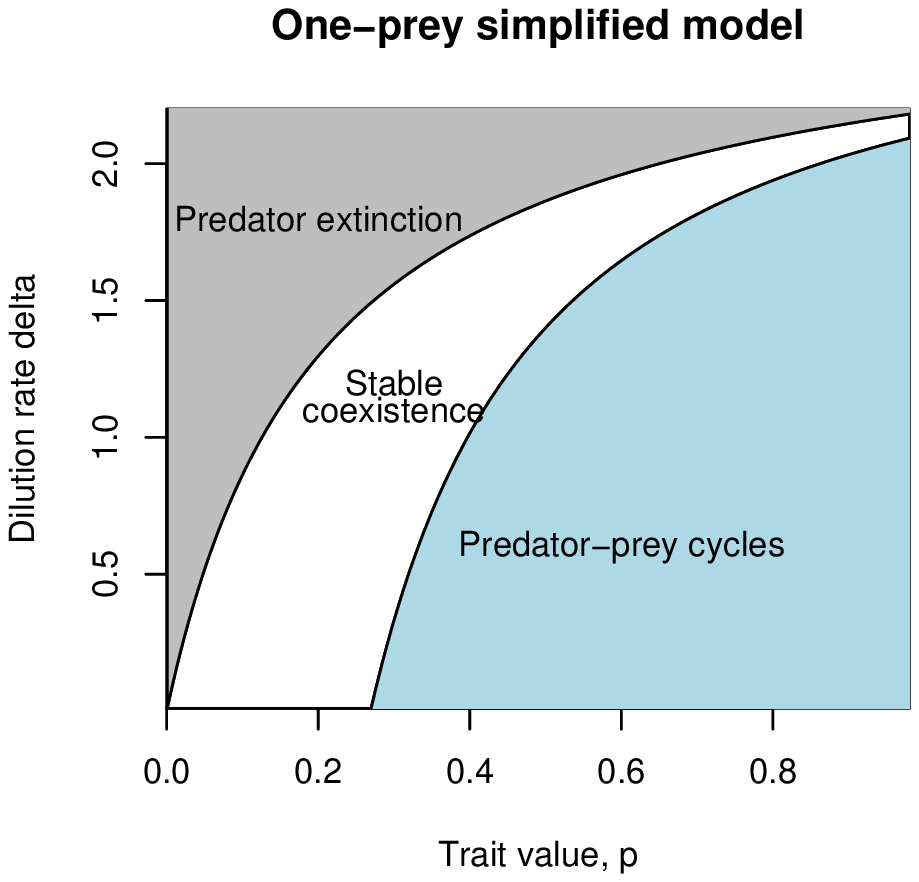}}
\caption{Bifurcation diagram for the rescaled, reduced clonal model with one prey type.}.  
\label{SingleCloneBifurcations}
\end{figure}

%\be
%\label{predatorSS}
%y_i = \bar \sigma_i -\bar x_i,\quad \mbox{where} \quad \bar\sigma_i = 
%(\bar x_i + \bar y_i) = \frac{1}{2}\left[\gamma_i - \sqrt{(\gamma_i^2 - 4m\bar x_i)}\right]
%\ee
%with $\gamma_i=k_i + 1 + m\bar x_i$, and $x_i = \frac{k_b}{p_i(g-1)}$.
We can use (\ref{coexistence2}) to derive the condition for predator-prey coexistence
in terms of the prey defense trait $p$ and the dilution rate $\delta$, recalling that
$\Lambda$ and $x_c$ are both implicit functions of $\delta$. Using  
\eqref{Lambda} and \eqref{xc} we obtain from \eqref{coexistence2}
\be
\frac{k}{m(\delta)-1} + \frac{k_b}{pg(\delta)-1} < 1, \quad \textrm{or} \quad p > 
\frac{1}{1-\Lambda}\left(\frac{k_b}{g(\delta)-1}\right).
\label{coexistence3}
\ee
\noindent
The quantity within parenthesis above is the amount of substrate present in perfect food (undefended prey with $p=1$).  
Solving \eqref{coexistence3} for $\delta$ in terms of $p$ yields the boundary between predator extinction and stable coexistence in Figure \ref{SingleCloneBifurcations}. To the left of this line, the predator goes extinct and the equilibrium is $E_0$. As the left-hand side of the second
expression in \eqref{coexistence3} is an increasing function of $p$, 
and the right-hand side is an increasing function of $\delta$, the range of $p$ values yielding  coexistence
narrows as $\delta$ increases (see Figure \ref{SingleCloneBifurcations}).

As in the standard Rosenzweig-MacArthur predator-prey model, the stability condition has
a graphical interpretation in terms of the nullclines. 
The prey nullcline is a parabola which peaks at 
$$ x^* = \frac{1}{m}\left[ 1-k + \sqrt{\Lambda}(m-2)\right].$$
The coexistence equilibrium is locally unstable if the peak of the prey nullcline is to the 
right of the predator nullcline (i.e., if $x^* > \bar x_c $). Note that a system with defended prey ($p < 1$)
is always more stable than a system with fully vulnerable prey ($p=1$)
as reductions in $p$ shift the predator nullcline to the right.

From \eqref{GeneralJacobian} the Jacobian of \eqref{eq7} at $E_c$ has the form
\be
J_c = \begin{bmatrix}
\bar x_c \frac{\partial \bar r_1}{\partial x}  & - \\
 + & 0 \\
\end{bmatrix} 
\label{Jc}
\ee
so $E_c$ is locally stable if the trace 
{\bf Tr}$(J_c)= \bar x_c \frac{\partial \bar r_1}{\partial x}$ is negative. Cycles emerge through a Hopf bifurcation
when the trace becomes positive. The condition {\bf Tr}$(J_c) \ge 0$ is equivalent to the following 
expression for model \eqref{eq7}:
\be
\frac{m k}{(k + 1 - \bar x_c- \bar y_c)^2}  \ge \frac{g p^2 \bar y_c}{(k_b + p \bar x_c)^2}
\label{OneCloneStability}
\ee
\cite{SmithWaltman1995}. Cycles begin when the rates of change 
in  prey substrate uptake (LHS) and in predator consumption (RHS) 
with respect to the amount of substrate present as prey ($x$) are exactly equal. 
Numerically solving \eqref{OneCloneStability} for $\delta$ in terms of $p$ yields the boundary 
between stable coexistence and predator-prey cycles in Figure \ref{SingleCloneBifurcations}. 
It is known that these cycles are stable and unique near the Hopf bifurcation, and numerical evidence uniformly 
indicates that they are always unique and attract all interior initial conditions except $E_c$ \cite{SmithWaltman1995}.

\subsection{Stability and dynamics of a two-prey system}
System \eqref{eqs6} has two prey types ordered so that 
$0 < p_1 < p_2 = 1$. We refer to prey 1 as the {\it defended} type and prey 2 as the {\it vulnerable} type. 
The cost for defense comes in the form of reduced growth rate, $1/k_1 \le 1/k_2$.  

Following Abrams \cite{Abrams1999}, we begin by finding the conditions for existence of an 
equilibrium $(\tilde{x}_1, \tilde{x}_2, \tilde y)$ at which all three population densities are positive;
we refer to this as a \textit{coexistence equilibrium}.  
Setting \eqref{eqs6} to zero and solving gives expressions for $\tilde X, \tilde Q$ and $\tilde y$ in
terms of model parameters (see Appendix \ref{SSAppendix}; as above $Q = p_1 x_1 + p_2 x_2$ is  
the total prey quality, and $X=x_1 + x_2$ is the total prey density). 
The prey steady states $\tilde x_1, \tilde x_2$ are then
\be
\left[ {\begin{array}{*{20}c}
\tilde x_1   \\
\tilde x_2   \\
\end{array} } \right] = \frac{1}{p_2  - p_1 }
\left[ {\begin{array}{*{20}c}
p_2 \tilde X - \tilde Q  \\
\tilde Q - p_1 \tilde X  \\
\end{array} } \right].
\label{xbars}
\ee
where 
\be
\tilde Q = \frac{k_b}{g-1}.
\label{Qbar}
\ee
A coexistence equilibrium thus exists provided $\tilde X >0 $ and $p_1 \tilde X < \tilde Q < p_2 \tilde X$,
or
\be 
\label{InvasionCondition}
p_1 < \frac{\tilde Q}{\tilde X} < p_2.
\ee

\begin{figure}[t]
\centerline{\includegraphics[width=5.5in]{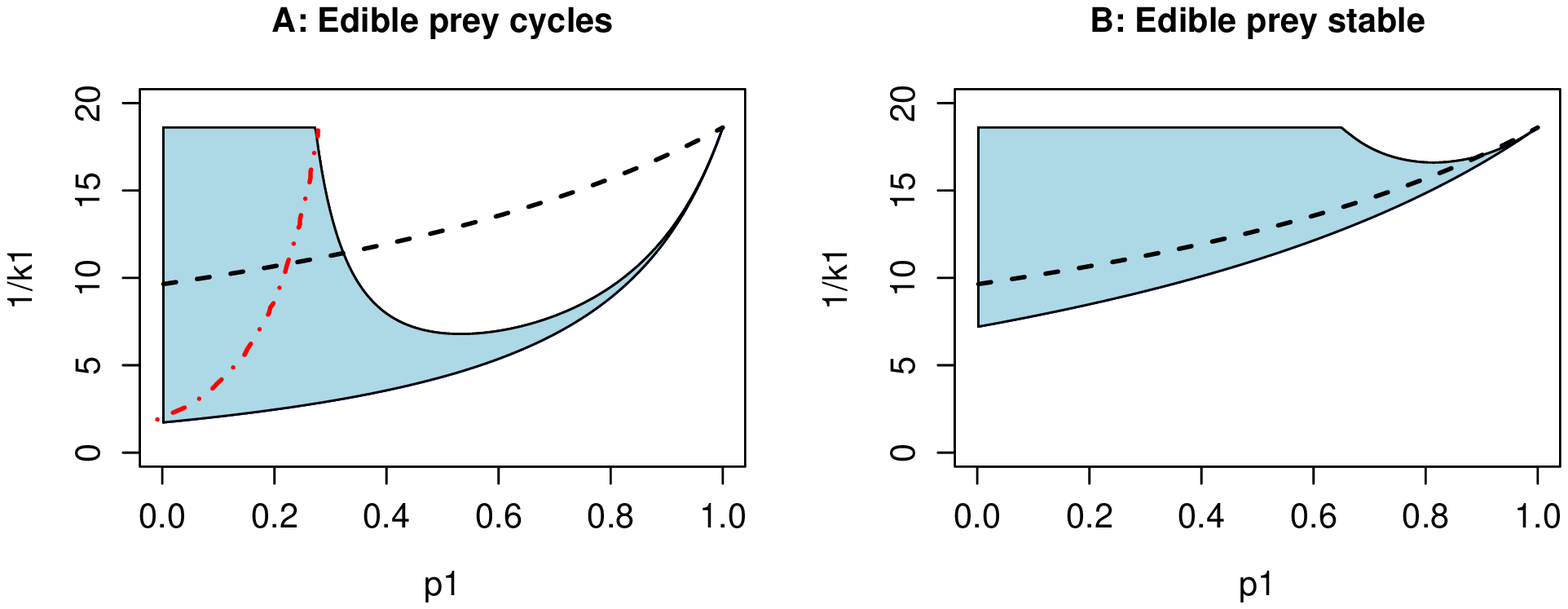}}
\caption{\small{Prey coexistence equilibria. The shaded gray regions indicate parameters $(p_1,1/k_1)$ for prey 
type 1 giving a coexistence equilibrium (stable or unstable) with the vulnerable prey type $p_2=1$. At $\delta=1.5$ (panel A) prey type 2 cycles, and at $\delta$ = 2.0 (panel B), prey type 2 is stable. The dashed lines
show a representative tradeoff curve \eqref{tradeoff}, assuming roughly 50\% reduction in growth rate as the cost of being 100\% defended. Here $k_c=0.017$, $a_1=1.0$, and $a_2=0.0165$.
In panel A the dash-dotted line indicates where the defended type can invade the limit cycle of the predator and vulnerable prey type.}}
\label{CCE}
\end{figure}

Beyond the above, system \eqref{eqs6} does not yield tidy 
analytical solutions for the steady states at coexistence. To study how parameter 
variation affects coexistence, we start by graphically mapping the region where a   
coexistence equilibrium exists as a function of the defended clone's parameters, $p_1$ and $k_1$ 
(Figure \ref{CCE}), without regard to whether or not the equilibrium
is locally stable. The coexistence region also varies with 
$\delta$, but selecting several $\delta$ values of interest gives
a general sense of how the coexistence region varies as a function of dilution rate. 

The lower boundary of the coexistence region occurs when the cost of
defense is so high that the equilibrium density of the defended prey $x_1$ drops 
to zero while $x_2$ and $y$ remain positive. Recalling the general form 
\eqref{GeneralForm}, the lower boundary is thus defined by the conditions 
$$
r_1(0,x_2,y)=r_2(0,x_2,y)=r_3(0,x_2,y)=0 \mbox{  with } x_2>0, y>0.$$ 
The conditions on $r_2$ and $r_3$ are solved by the steady state $E_2 = (0, \bar x_2, \bar y_2)$
for a one-prey system with only the vulnerable prey. 
The lower boundary of the coexistence region is thus defined by the condition 
$r_1(0,\bar{x}_2,\bar{y}_2)=0$, which can be written as 
\be
\label{Lowercondition}
\frac{1}{k_1}=\frac{1}{1 -\bar x_2 - \bar y_2}\left[\frac{\bar y_2p_1 + \bar x_2}{\bar x_2(m-1)- \bar y_2 p_1}\right].
\ee

The upper boundary of the coexistence region occurs when the cost of defense is so low
that the defended prey (at the equilibrium density) drives one of the other populations to extinction. 
In section \ref{GeneralEvoCycles} we show that for $p_1 < p^*=\tilde Q/(1-\Lambda_2)$, 
the predator goes extinct first ($\tilde y \to 0$) as $k_1$ decreases, because the defended prey (at steady state) 
drives the vulnerable prey to low abundance and the defended prey is very poor food. This occurs 
at $k_1 = k_2$ (zero cost of defense). For $p_1 > p^*$, the vulnerable prey type 
is outcompeted by the defended type before $k_1$ has reached $k_2$. 
This boundary is therefore defined by the conditions 
$$
r_1(x_1,0,y)=r_2(x_1,0,y)=r_3(x_1,0,y)=0 \mbox{ with } x_1>0,y>0. 
$$
\noindent
The conditions on $r_1$ and $r_3$ are solved by the one-prey steady state $E_1 = (\bar x_1, 0, \bar y_1)$, so 
the condition $r_2(\bar{x_1},0,\bar{y_1})=0$ defines the upper boundary of the coexistence region
for $p>p^*$. The upper boundary of the coexistence region is thus the curve  
\be
\frac{1}{k_1}= \textrm{min}\left[\frac{1}{k_2}, \frac{1}{\varphi(\bar x_1,\bar y_1)} \right]
\label{UL}
\ee
\noindent
where $\varphi$ is value of $k_1$ that solves 
\be
\label{UpperCondition}
\frac{m (1 - \bar x_1 - \bar y_1)}{k_2 + (1-\bar x_1-\bar y_1)}-\frac{\bar y_1 + p_1 \bar x_1}{(p_1 \bar x_1)}  =  0,
\ee
noting that $\bar x_1$ and $\bar y_1$ are functions of $k_1$ and $p_1$. 
The two segments of the upper boundary defined by \eqref{UL} meet at the point
$$ p_1=p^*= \frac{\bar Q}{1-\Lambda_2}, \quad k_1=k_2.$$ As $\delta \to 0$ (with the
parameter scalings in Table 2), $\bar Q \downarrow 0$ and $\Lambda_2 \uparrow 1$, so $p^* \uparrow 0$  
so there is typically an increasingly narrow band of $p_1$ values for which a $p_1 -- k_1$ tradeoff curve
lies in the coexistence equilibrium region (unless the tradeoff curve happens to lie exactly inside
the cusp of the coexistence equilibrium region).   

As $p_1 \rightarrow 1$, the upper and lower boundaries of the coexistence region
meet at $p_1=p_2=1, k_1=k_2$ (Figure \ref{CCE}). That is, if the two prey are almost equally vulnerable
to predation, they can only coexist at equilibrium if a tiny bit of defense has a tiny cost.   
To prove that this occurs, we show that the point $p_1=1,k_1=k_2$ lies on both boundaries. 
At this point the two prey are identical so $\bar x_1 =  \bar{x_2}$ and $\bar y_1 = \bar y_2$.
The upper boundary is defined by $r_2(\bar x_1,0,\bar y_1)=0$.
At $p_1=1, k_1=k_2$, 
$$
r_2(\bar x_1,0,\bar y_1)=r_2(\bar x_2,0,\bar y_2)=r_2(0,\bar x_2,\bar y_2)=0,
$$
thus $p_1=1, k_1=k_2$ lies on the upper boundary.
The lower boundary is defined by $r_1(0,\bar x_2,\bar y_2)=0$.
At $p_1=1, k_1=k_2)$, 
$$
r_1(0,\bar x_2,\bar y_2)=r_1(0,\bar x_1 \bar y_1)=r_2(\bar x_1,0,\bar y_1)=0,
$$
which shows that $p_1=1, k_1=k_2$ lies on the lower boundary. 
Thus both boundaries converge to $k_1=k_2$ as $p_1 \to 1$. 

\subsection{Local stability of coexistence equilibria}
\begin{figure}[t]
\centerline{\includegraphics[width=6in,height=6in]{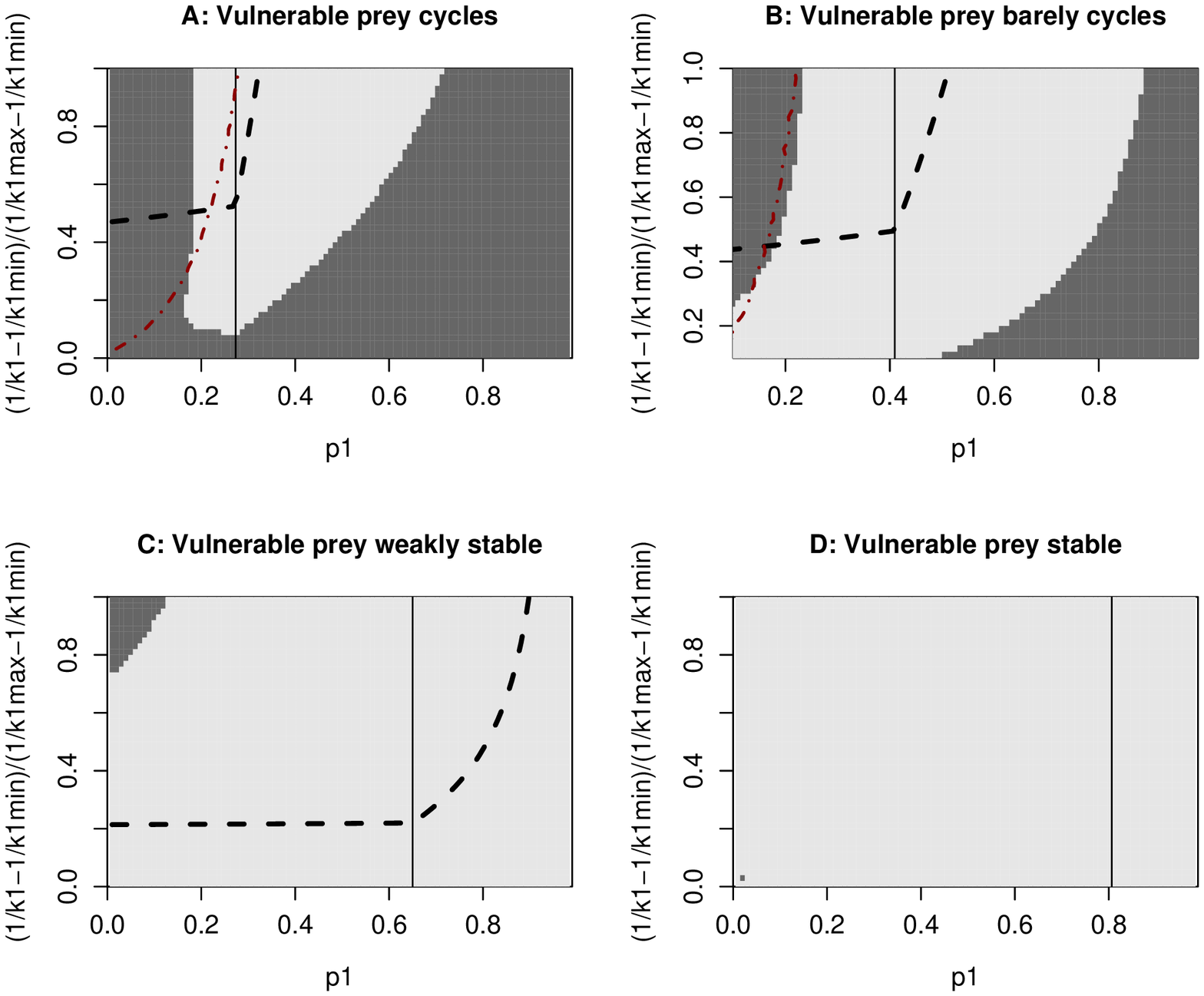}}
\caption{\small{Stability of coexistence equilibria for the reduced two-prey model. 
In each panel, the horizontal axis is the palatability $p_1$ of the defended prey with the model 
scaled so that $p_2=1$. To remain consistent with Abrams ({\it cf.} \cite{Abrams1999}, Figures 1 - 3) the vertical axis is $1/k_1$, scaled so that 0 and 1 correspond to the lower and upper limits of the coexistence equilibrium region (Figure \ref{CCE}). 
The Jacobian matrix and its eigenvalues were 
evaluated at an even $50 \times 100$ grid of values. Lighter gray indicates that the equilibrium is stable, darker gray that it is unstable; in all cases the computed eigenvalues with largest real part are a complex conjugate pair. 
The vertical black line is at $p_1=p^*$, the value where the straight and curved segments of the upper limit of the
coexistence equilibrium region meet.  The dashed curve in panels A and B is the tradeoff curve 
$k_1=k_c + \alpha_2(1-p_1)^{\alpha_1}$, with $k_c=k_2=0.054,
\alpha_1=1; \alpha_2=0.05$ at $p_1$ values lying within the coexistence equilibrium region. 
The dash-dotted line represents the 
minimum $1/k_1$ values at which the defended prey can invade the (predator + vulnerable prey) limit cycle 
(see Appendix \ref{InvadeEdibleCycle}). Parameter values for these plots are as follows:  panel  A., $\delta=1.5$; B., $\delta=1.75$; C., $\delta=2.0$ and D., $\delta=2.1$. All other parameters are unchanged and are as shown in Table 2.
}}
\label{CCEStability}
\end{figure}

To characterize two-prey evolutionary cycles we need to find the bifurcation 
curves in parameter space where these cycles arise. 
The ``empirical facts'' are summarized in Figure \ref{CCEStability}, based on
numerical evaluations of the Jacobian and its eigenvalues within the coexistence equilibrium region.
In Figure \ref{CCEStability} we change the stability of the {(predator + vulnerable prey)} system by varying 
the value of $\delta$, but the results are qualitatively the same if other parameters are varied instead 
(e.g., varying the prey maximum growth rates).  

The stability properties in Figure \ref{CCEStability} explain the major qualitative
features of the two-prey model's bifurcation diagram (Figure \ref{DynamicsFigs}D). To see the
connection, recall that a horizontal (constant $\delta$) slice through Fig. \ref{DynamicsFigs}D
corresponds to a tradeoff curve between $p_1$ and $k_1$ in the panel of Figure \ref{CCEStability}   
with the same value of $\delta$. Panel A of Figure \ref{CCEStability} has $\delta=1.5$. 
When $p_1$ is near 1, the tradeoff curve lies above the coexistence equilibrium region,
and the defended prey type eventually outcompetes the vulnerable type. For $p_1$ very close to 1 
the prey types are very similar, and the vulnerable type persists for a long time. 
The system exhibits ``classical'' predator-prey cycles as if a single prey-type were present, even though two types are transiently coexisting. 
For $p_1$ somewhat smaller, the vulnerable type is quickly eliminated and there are either 
classical cycles with only the defended type (open circles in Fig. \ref{DynamicsFigs}D),
 or (for lower values of $p_1$) the defended prey type goes to a stable equilibrium with the
predator (open triangles in Fig. \ref{DynamicsFigs}D).
As $p_1$ decreases further, Figure \ref{CCEStability}A shows that the tradeoff curve
enters the coexistence region in the area where the coexistence equilibrium is stable, so
the system then exhibits stable coexistence (cross-hatching in Fig. \ref{DynamicsFigs}D). Finally,
as $p_1$ decreases towards 0, the tradeoff curve enters the area where the coexistence equilibrium
is unstable, and it lies above the dash-dot curve marking the $k_1$ value required for the defended prey type to invade the
vulnerable prey's limit cycle with the predator. The system
exhibits evolutionary cycles with both prey types persisting (filled circles at $p \approx 0$ in Fig. \ref{DynamicsFigs}D).   

Figure \ref{CCEStability}A also shows that there is a region of parameters (below the dash-dot curve) where the coexistence equilibrium is stable and the system therefore has coexisting attractors: a locally
stable coexistence equilibrium, and a locally stable limit cycle with the predator and the vulnerable prey. 

Figure \ref{CCEStability}B, which has $\delta=1.75$, shows the same sequence of transitions  
as Figure \ref{CCEStability}A, but each occurs at higher values of $p_1$, reflecting the stabilizing effect of increased
washout. This is reflected in Figure \ref{DynamicsFigs}D: increasing $\delta$ above 1.0 shifts 
all the bifurcation boundaries to higher $p$ values, but the sequence of bifurcations as 
$p$ decreases is unchanged. However for $\delta$ sufficiently high (panels C and D in Figure \ref{CCEStability}),
the tradeoff curve lies either below the coexistence equilibrium region or within the region where
the coexistence equilibrium is stable, so evolutionary cycles never occur. Instead, there is either
stable coexistence of the two prey with the predator, or classical predator-prey cycles with only
the vulnerable prey type. 

Evolutionary cycles are also eliminated -- but for a different reason -- 
as $\delta \downarrow 0$ in Figure \ref{DynamicsFigs}D. As noted above, as $\delta \downarrow 0$ we also
have $p^* \downarrow 0$, so unless $p_1 \approx 0$ the tradeoff curve lies above
the coexistence equilibrium region and only the defended prey persists with the predator, 
cycling at higher $p_1$ and stable at lower $p_1$. Only very near $p_1 =0$, 
a region tiny enough to be missed by our simulation grid in Figure \ref{DynamicsFigs}, can
there be coexistence of both prey with the predator. 

\paragraph{Stability on the edges.} We can gain some understanding of the patterns in Figure \ref{CCEStability}, and
see that they are not specific to the parameter values used to draw the Figure, by examining the limiting cases
that occur along the edges of the coexistence equilibrium region. One general 
conclusion (explained below) is that the bottom and right edges, and the right-hand portion 
of the top edge, all must have the same stability as the reduced system with the predator
and only the vulnerable prey (prey type 2). However even if this system is unstable, there
must be a region along the top edge where the coexistence equilibrium is stable. 

The Jacobian matrix that determines equilibrium stability is derived in 
Appendix \ref{JacobianAppendix}. We also show there that the determinant of this Jacobian is always negative 
at a coexistence equilibrium unless $p_1=p_2$, so the coefficient $c_0 = -\det(J)$ in the Routh-Hurwitz stability 
criterion for 3-dimensional systems is always positive. 

\paragraph{Bottom and right edges:} Near the bottom and right edges, the coexistence equilibrium
has the same local stability as the {(predator + vulnerable prey)} subsystem (panels A and B versus
C and D in Fig \ref{CCEStability}). The bottom edge is the lower limit of the
coexistence equilibrium region, where $\tilde x_1 \to 0.$ The coefficients for the Routh-Hurwitz stability criterion (see Appendix \ref{HopfConditions}) are then  
\be
c_0 =-\det(J) > 0, \quad c_1 = T_2(J) \to \delta_2, \quad c_2 = -T(J) \to -\tau_2 
\ee
where $\delta_2$ and $\tau_2$ are the determinant and trace, respectively, of the
$2 \times 2$ Jacobian for the (predator + vulnerable prey) system. If this 
one-prey system is stable then $\delta_2 > 0, \tau_2 <0$ so $c_0, c_1$ and $c_2$
are all positive. Moreover $c_0=\emph{O}(\tilde x_1)$ (see Appendix \ref{JacobianAppendix}), 
so when $\tilde x_1$ is small we have $c_1 c_2 > c_0$ and the equilibrium is stable. Conversely if the steady state for the {(predator + vulnerable prey)} system is 
unstable, $c_2$ is negative so the full system is also unstable. 

The right edge corresponds to the cusp in the coexistence region as $p_1 \to 1$. 
Near the cusp the two prey become increasingly similar ($p_1 \approx p_2=1, k_1 \approx k_2$). 
Using \eqref{GeneralJacobian}, the functional forms of the $r_i$ and the fact that
$p_1 \approx p_2$ then imply that the form of $J$ is approximately  
\be
J_0  = \left[ {\begin{array}{*{20}c}
   {a q} & {a q} & { - qb}  \\
   {a (1 - q)} & {
a (1 - q)} & { - (1 - q)b}  \\
   c & c & 0  \\
 \end{array} } \right]
\label{Jclose}
\ee
where $q=\tilde x_1/(\tilde x_1 + \tilde x_2)$; even if $p_1$ is near $p_2$, it is not necessarily the
case that $\tilde x_1$ is close to $\tilde x_2$. In \eqref{Jclose} $b$ and $c$ are positive while 
$a$ has the sign of $\partial r_1/\partial x_1$ which may be positive or negative. 
One eigenvalue of $J_0$ is 0, corresponding to the dynamics of $x_1-x_2$. The others are 
$\frac{1}{2}(a \pm \sqrt{a^2 - 4bc}),$ 
which are also the eigenvalues of a single-prey system at the coexistence steady state. Thus, the two-prey system 
with $p_1 \approx p_2=1$ ``inherits'' two eigenvalues from the one-prey system with $p = 1$. 
 
When the one-prey system with p=1 is cyclic, the inherited eigenvalues are a 
complex conjugate pair. In the corresponding eigenvectors, the components for the two clones are identical when
$p_1 = p_2$. This implies that when $p_1 \approx p_2$ the eigenvector 
components will be similar, so the two prey types cycle almost exactly in phase. 
The period of these oscillations is determined by the inherited eigenvalues, 
so it is close to the period of the corresponding one-prey system. 

When the one-prey system is stable, the Routh-Hurwitz criterion 
(Appendix \ref{HopfConditions}), using $J_0$ to approximate trace$(J)$ and $T_2(J)$ and the 
fact that $\textrm{det}(J) < 0$ for $p_1 \not= p_2$ , implies that the full system will also be stable.
Therefore, a coexistence equilibrium with two nearly identical prey   
has the same stability as the equilibrium for the corresponding one-prey systems. During damped
oscillations onto a stable coexistence equilibrium, or diverging oscillations away from 
an unstable one, the clones will oscillate nearly in phase with each other and inherit 
the cycle period of the one-prey system. 

\paragraph{Top edge:} The rightmost portion of the top edge also corresponds to the cusp in the coexistence
equilibrium region, so the stability here is also the same as that of the 
{(predator + vulnerable prey)} system. In general, as $1/k_1$ approaches the upper limit of 
the coexistence equilibrium region when $p_1> p^*$ (the curved portion), the stability of the 
two-prey system approaches that of the {(predator + defended prey)} system with $1/k_1$ approaching $1/k_2$.  
This must be stable if the {(predator + vulnerable prey)} system is stable, because the defended
prey is always more stable, as noted above. If the {(predator + vulnerable prey)} system cycles, 
then there will be instability as $p_1 \to 1$ along the top edge. 

However, there is always stability near the top edge
for $p_1 \to p^*$, as follows. Along the straight portion of the top edge ($p_1 < p^*$),
as $1/k_1$ approaches the edge, the coexistence equilibrium converges to a limit with $\tilde y=0$, 
while along the curved portion the limiting coexistence equilibrium has $\tilde x_2=0$. So near their intersection at
$p_1=p^*$, both $\tilde x_2$ and $\tilde y$ approach 0. Condition \eqref{OneCloneStability}
then implies that the {(predator + defended prey)} system is stable, so the coexistence equilibrium is
stable near the top edge for $p_1$ just above $p^*$. By continuity, there is an open region of
$(p_1,k_1)$ values near $p_1=p^*,k_1=k_2$ where the coexistence equilibrium is locally stable.  
If the {(predator + vulnerable prey)} system is only weakly unstable then this 
stability region may be quite large (Fig \ref{CCEStability} panel B), but it cannot 
reach either the bottom or right edges. 

\paragraph{Left edge:} Finally, consider the edge $p_1 = 0$. The steady states simplify to  
\be
\tilde x_1=1- \tilde Z- \tilde x_2 - \tilde y, \quad \tilde x_2 = \tilde Q = \frac{k_b}{g-1}, 
\quad \tilde y= \tilde Q(m-1)\frac{(k_1 - k_2)}{k_1 + k_2(m-1)}
\label{leftedgeSS}
\ee
where $\tilde Z = \frac{k_1}{m-1}$. The coexistence equilibrium exists for $\vartheta < \frac{1}{k_1} < \frac{1}{k_2}$
where $\vartheta$ is the value of $1/k_1$ that solves $\tilde x_2 + \tilde y + \tilde Z=1$, 
noting that $\tilde y$ depends on $k_1$. The Jacobian matrix at \eqref{leftedgeSS} is 
\be
\begin{gathered}
  J = \left[ {\begin{array}{*{20}c}
   { -a_1\tilde x_1  } & { -a_1 \tilde x_1 } & {-a_1\tilde x_1 }  \\
   { -a_2\tilde x_2 } & { (-a_2 + g\tilde y F^2)\tilde x_2 } & {-(a_2 + gF)\tilde x_2 }  \\
   {0} & {gk_b\tilde y F^2} & 0  \\
 \end{array} } \right] \\ 
\end{gathered}
\ee
where setting $p_1 = 0$ and $p_2 = 1$ gives $F=\frac{1}{k_b + \tilde x_2}$ and 
\be 
a_i= \frac{m k_i}{(k_i + \tilde{Z})^2}. 
\label{defnAi}
\ee
Near the lower limit of the left edge, we know that the system inherits the stability of the 
{(predator + vulnerable prey)} system. Above the lower limit we can use the 
Routh-Hurwitz criterion (Appendix \ref{HopfConditions}) to determine stability.  
The coefficients $c_0$ and $c_1$ have common factor  $\tilde{x}_2gF^2 > 0$. Dividing this out 
gives modified coefficients
\be
\tilde c_0 = a_1\tilde{x}_1gk_bF > 0, \quad \tilde c_1 = k_b(a_2 + gF) - a_1\tilde x_1, 
\quad \tilde c_2 = a_1\tilde x_1 + \tilde{x}_2(a_2 - g\tilde{y}F^2)
\ee
and the stability conditions remain the same:
$
\tilde c_0, \mbox{ }\tilde c_1, \mbox{ }\tilde c_2 > 0, \mbox{ }\tilde c_1\tilde c_2 > \tilde c_0.
$
Extensive numerical evaluations of the coefficients as $\delta$ is varied indicate that 
loss of stability occurs when the condition $\tilde c_1\tilde c_2 - \tilde c_0 > 0$  is violated -- 
the equilibrium is stable if this condition holds and unstable if it fails. Assuming this is true, 
loss of stability along the left edge occurs via a Hopf bifurcation (Appendix \ref{HopfConditions}).  

\begin{figure}[t]
\centerline{\includegraphics[width=4.25in]{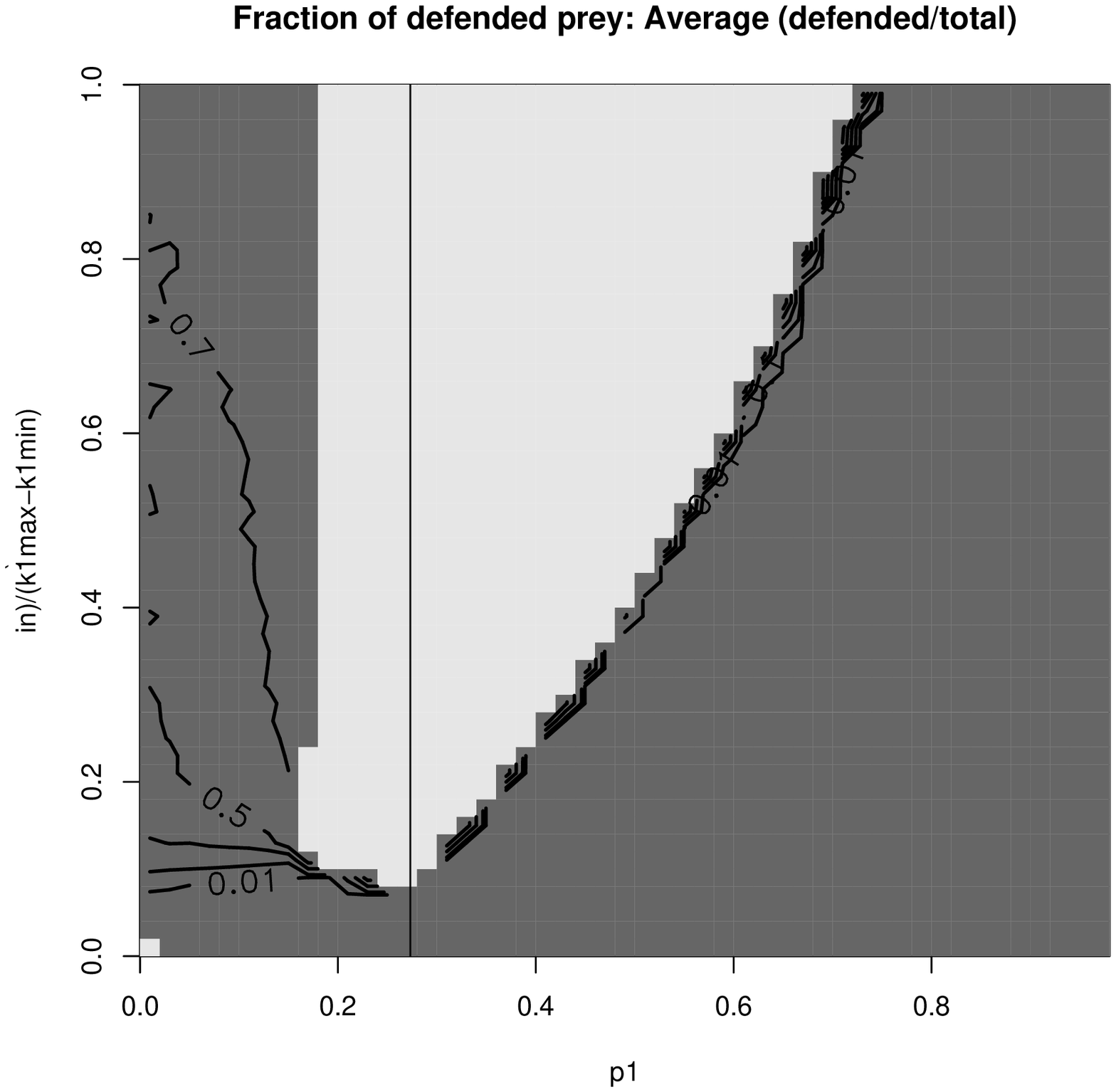}}
\caption{\small{Contour plot of the long-term average fraction of defended prey. The horizontal axis is the
palatability $p_1$ of the defended prey, with the model scaled so that $p_2=1$. The vertical axis represents $1/k_1$, with 0 and 1 corresponding to the lower and upper limits
of the coexistence equilibrium region (Figure \ref{CCE}). Numerical solutions of the
model were used to compute the long-term average value of $x_1/(x_1+x_2)$ for
parameter values such that the (predator + vulnerable prey) system (same parameter values
as panel A of Fig. \ref{CCEStability}). In the lighter-gray region
the coexistence equilibrium is stable. In the darker-gray region the equilibrium is unstable. The vertical black
line is at $p_1=p^*$, the value where the straight and curved segments of the upper limit of the
coexistence equilibrium region meet. 
%The dash-dot line is the minimum $1/k_1$ value at
%which the defended prey can invade the {(predator + vulnerable prey)} limit cycle. 
Parameter values are as in Table 2 with $\delta=1.5$. 
}}
\label{ContourFractionDefended}
\end{figure}

\subsection{The structure of evolutionary cycles}
The stability analysis above delimits the situations in which evolutionary cycles occur. 
As illustrated in Figure \ref{CCEStability}A, they arise when the $p_1$ versus $k_1$ tradeoff 
curve passes (with decreasing $p_1$) from the region of stable coexistence equilibria 
near $p_1=p^*, k_1=k_2$ to the region of unstable coexistence equilibria with $p_1\approx 0$.
For $1/k_1$ below the dash-dot curve in Figure \ref{CCEStability}A the defended prey cannot invade
the vulnerable prey-predator limit cycle (see Figure \ref{ContourFractionDefended}). As $1/k_1$ increases,
the defended prey becomes persistent and then increases in average abundance. As $1/k_1 \to 1/k_2$ the
characteristic features of evolutionary cycles emerge: longer cycle period and out-of-phase oscillations
in predator and total prey abundance. 

\begin{figure}[t]
\centerline{\includegraphics[width=5.5in]{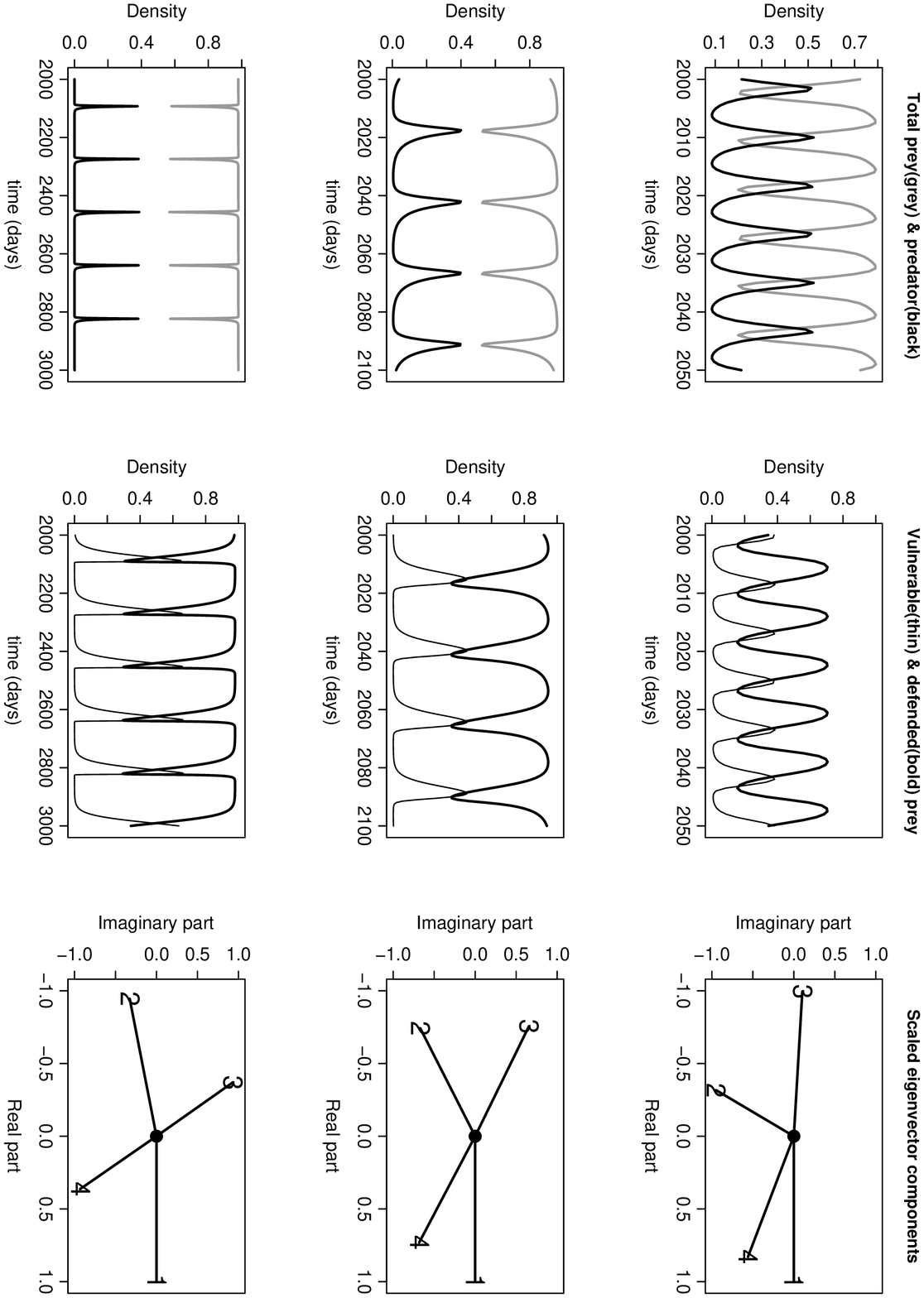}}
\caption{\small{Coexistence of edible and defended prey on a limit cycle. 
Parameter values for all plots were $\delta=0.9, m=3.3/\delta, g=2.3/\delta,
k_2=0.05, k_b=0.2, p_2=1, p_1=0.08$. Values of $k_1$ were 0.4 (top row),
0.1 (center row) and 0.055 (bottom row). In each row the leftmost panel
shows the dynamics of total prey and predator densities, the center panel
shows the dynamics of the two prey types, and the rightmost panel shows the phases
of the Jacobian matrix components: 1=defended prey, 2=edible prey, 3=predator,
4=total prey. }}
\label{CoCycles}
\end{figure}

The phase relations on evolutionary cycles can be seen in the dominant eigenvector of the Jacobian matrix for the unstable fixed point (Figure \ref{CoCycles}). There is a codominant pair
of complex conjugate eigenvalues, and ( because $\det(J)<0$ ) the third eigenvalue is real and
negative. As the defended prey has very low palatability, the predator and 
the vulnerable prey have the classical quarter-phase lag. Here the phase angle is $90^{o}$; because eigenvectors are only defined up to arbitrary scalar multiples, including arbitrary
rotations in the complex plane from multiplication by $e^{i \theta}$, only the relative phases
of eigenvector components are meaningful.  As $1/k_1$ increases, the eigenvector components for the two prey types become out of phase with each other ($\approx 180^{o}$ phase angle, right column of Figure \ref{CoCycles}). As a result, the predator and total prey densities are out of phase with each other. 

In the next section we show that these phase relations become
exact as the limit $k_1 \to k_2$ is approached, for a general version of the model 
in which we do not specify the functional forms of the predator and prey functional responses. 

\section{Evolutionary cycles in a general two-prey model} \label{GeneralEvoCycles}
In this section we analyze the limiting phase relations in evolutionary cycles, 
for $p_1 \ll 1$ and low cost to defense, without specifying the functional forms
of the prey and predator functional responses. We consider a two-prey, one-predator model 
that (after rescaling) can be written in the form 
\be
\begin{gathered}
\dot x_i  = x_i \left( f(X,y,\theta_i) - p_i y g(Q) \right),i = 1,2  \hfill \\
\dot y = y\left(Qg(Q) - d \right) \hfill \\ 
\label{pmc}
\end{gathered} 
\ee
where as usual $X=x_1+x_2$ is the total density of prey and $Q=p_1 x_1 + p_2 x_2$ is the
total prey quality as perceived by the predator. 
The key assumption in \eqref{pmc} is total niche overlap in the prey types (e.g., because they
are two clones within a single species), which is reflected
in $f$ being a function of $X$. To model the trophic relations, $f$
is assumed to be strictly decreasing in $X$ and nonincreasing in $y$, and $h(Q)=Qg(Q)$ is strictly 
increasing in $Q$. Model \eqref{pmc} includes the Abrams-Matsuda model 
\cite{AbramsMatsuda1997} (a two-prey version of the 
Rosenzweig-MacArthur model with Lotka-Volterra competition between the prey) and the two-prey, one-predator chemostat model analyzed in the previous section of this paper. 

The parameter $\theta_i$ in \eqref{pmc} represents the prey-specific cost of defense, 
with $f$ decreasing in $\theta$. Because $f$ is decreasing in $X$, we can parameterize $f$ 
such that $\theta$ is the steady-state density for a single prey-type in the absence of predators, i.e. 
\be
f(\theta,0,\theta) = 0. 
\label{deftheta}
\ee  
As usual we number the prey so that $p_1 < p_2$ and therefore $\theta_1 < \theta_2$, and 
rescale the model so that $p_2=1$. 

Evolutionary cycles can be viewed as arising when $\theta_1 \uparrow \theta_2$ with $p_1 \ll 1$, such that when $\theta_1 \approx \theta_2$ 
\begin{enumerate}
\item There is a positive coexistence equilibrium with $\tilde x_1, \tilde x_2$ converging to
positive limits while $\tilde y \rightarrow 0$ as $\theta_1 \uparrow \theta_2$;
\item The coexistence equilibrium is an unstable spiral for  $\theta_1 \approx \theta_2$. 
\end{enumerate}
At the end of this section, we show that Condition 1 always holds for \eqref{pmc}, and in
Appendix \ref{ECeigenvals} we show that the coexistence equilibrium is always a spiral.
To determine the limiting phase relations, we need to find the eigenvector corresponding
to the dominant eigenvalue with positive imaginary part (see Appendix \ref{EigenPhase}): the relative phase
angles of this eigenvector's components (in the complex plane) correspond to the phase lags between
the corresponding state variables in solutions to the linearized system near the steady state. 

The Jacobian for \eqref{pmc} in the limit described above is  
\be
J_0 = \left[ {\begin{array}{*{20}c}
    \tilde x_1 \tilde f_X  &  \tilde x_1 \tilde f_X & \tilde x_1 f_y -p_1 \tilde x_1 \tilde g  \\
    \tilde x_2  \tilde f_X &  \tilde x_2 \tilde f_X & \tilde x_2 f_y -\tilde x_2 \tilde g  \\
   0 & 0 & 0  \\
\end{array} } \right]. 
\label{LimitJacobian}
\ee
The characteristic polynomial of \eqref{LimitJacobian} factors to show that
the eigenvalues of \eqref{LimitJacobian} are $f_X(\tilde x_1 + \tilde x_2)<0$ and 0 as a repeated root; the first has 
corresponding eigenvector $(\tilde x_1,\tilde x_2,0)$, and for 0 there is the unique eigenvector $(1,-1,0)$. 
The zero eigenvalue therefore has algebraic multiplicity 2 and geometric multiplicity 1.

The long period of evolutionary cycles is explained by the fact that the dominant
eigenvalues converge on a double-zero root as $\theta_1 \to \theta_2$. The cycle period near the
fixed point is inversely proportional to the imaginary part of the dominant 
eigenvalues, which converges to 0 in this limit. 

To determine the limiting phase relations consider
a small perturbation of the defended prey parameters, $\theta_1 = \theta_2-\epsilon$. 
For $\epsilon$ small, we show in Appendix \ref{ECeigenvals} that the double-zero 
eigenvalue is perturbed to a complex conjugate pair of eigenvalues. To study cycles,
we assume that these have positive real part. That is, near the double-zero root the (scaled) characteristic polynomial is
perturbed to leading order from $p(z)=z^2$ to $p(z)=(z-\epsilon a)^2+\epsilon b^2$ for some $b>0$, so the perturbed
eigenvalues have $\textit{O}(\epsilon)$ real part and imaginary parts $\pm \sqrt{\epsilon}bi$ to leading order
(here $i = \sqrt{-1}$).
We need to determine the corresponding perturbed eigenvectors. Let 
$\vec{w_0}$ denote the unperturbed eigenvector $(1,-1,0)$, and let $\vec{w_0} + \vec{w_e}$ be a perturbed eigenvector
corresponding to the complex eigenvalue with positive imaginary part,  
scaled so that its first component is 1. The first component of $\vec{w_e}$ is therefore 0. The perturbed
Jacobian is $J_0 + \epsilon J_1$ for some matrix $J_1$. Then  
\be
(J_0+ \epsilon J_1)(\vec{w_0}+\vec{w_e}) = \sqrt{\epsilon} b i (\vec{w_0}+\vec{w_e}) + \textit{O}(\epsilon).
\label{PerturbJ}
\ee 
Using $J_0 \vec{w_0}=0$ and keeping only leading-order terms, gives 
\be
J_0 \vec{w_e} = \sqrt{\epsilon} bi \vec{w_0}.
\label{PerturbJ}
\ee 
Let $\vec{w_e} = (0,w_2,w_3)$; then writing out \eqref{PerturbJ} in full, $w_2$ and $w_3$ satisfy
\be
\left[ {\begin{array}{*{20}c}
\tilde x_1 f_X   & \tilde x_1 f_y -p_1 \tilde x_1 \tilde g  \\
\tilde x_2 f_X   & \tilde x_2 f_y - \tilde x_2 \tilde g  \\
\end{array} } \right]
\left[ {\begin{array}{*{20}c}
w_2 \\
w_3 \\
 \end{array} } \right]
 = \sqrt \epsilon bi\left[{\begin{array}{*{20}c}
 1 \\
 -1\\
\end{array} } \right].
\label{w23}
\ee
$w_2$ and $w_3$ must be pure imaginary, because the unique solution to the real part
of \eqref{w23} is $(0,0)$. Writing $w_j=(\sqrt{\epsilon}bi)z_j$ and solving for the $z$'s,
we find that $z_2<0$ and $z_3>0$; specifically 
\be
\left[ {\begin{array}{*{20}c}
z_2   \\
z_3  \\
 \end{array} } \right] \propto \left[ {\begin{array}{*{20}c}
(\tilde x_1 + \tilde x_2) f_y - (p_1 \tilde x_1 + \tilde x_2) \tilde g  \\
-(\tilde x_1 + \tilde x_2) f_X \\
\end{array} } \right]
= \begin{bmatrix}
\tilde X f_y - d  \\
-\tilde X f_X \\
\end{bmatrix} 
\ee
using the fact that (from the second line of \eqref{pmc})
\be
\tilde Q g(\tilde Q) = d.
\label{defQbar}
\ee
So to leading order the eigenvector corresponding to 
eigenvalue $\sqrt{\epsilon} bi + o(\sqrt{\epsilon})$ is 
\be
\vec{w_0} + \vec{w_e} = 
\left[ {\begin{array}{*{20}c}
 1  \\
 - 1 - \sqrt{\epsilon}Bi  \\
 \sqrt{\epsilon} Ci  \\
 \end{array} } \right] \mbox{ for some } B,C >0. 
\label{PerturbVec}
\ee
Now add total prey as a fourth state variable to the system. The corresponding eigenvector 
component is the sum of the first two components in \eqref{PerturbVec} (see
Appendix \ref{EigenPhase}):    
\be
\left[ {\begin{array}{*{20}c}
 1  \\
- 1 - \sqrt{\epsilon} Bi  \\
\sqrt{\epsilon} Ci  \\
- \sqrt{\epsilon} Bi \\
 \end{array} } \right] 
\label{PhaseVec1}
\ee
We can multiply each component of \eqref{PerturbVec} by an arbitrary
real constant without affecting the phase angles, so we can consider instead
$\left(1, - 1 - \sqrt \epsilon  Bi, i,-i \right).$ Then as $\epsilon \to 0$ the vector 
giving the relative phases for prey 1, prey 2, predator, and total prey becomes 
\be
\left[ 1\quad -1\quad i\quad -i\right]^T.  
\label{PhaseVec2}
\ee
The components of the limiting phase-angle vector \eqref{PhaseVec2} lie exactly on the coordinate axes. 
The two prey types (first and second eigenvector components) are exactly out of phase; the predator and total prey (third
and fourth components) are exactly out of phase; and there is a quarter-period lag between
the vulnerable prey and the predator. This holds in the limit $\theta_1 \to \theta_2$ for $p_1 < p^*$ 
such that the coexistence equilibrium remains is an unstable spiral when $\theta_1 < \theta_2$. 

Finally, we now show that for $p_1$ sufficiently small there always exists coexistence 
equilibria as $\theta_1 \uparrow \theta_2$ such that $\tilde x_1,\tilde x_2$ approach finite 
limits while $\tilde y \to 0$. Evolutionary cycles then occur whenever these equilibria 
are unstable. 

We must have $\theta_2 > \tilde Q$ -- this is just the condition that the vulnerable prey, at steady state, is a sufficient
supply of food for the predator to increase when rare. \textit{Fix a value of} $\tilde y > 0$ -- we will show that for any $\tilde y$
sufficiently small, there will be a value of $\theta_1 \approx \theta_2$ such that there is a coexistence equilibrium 
with $\tilde y$ as the equilibrium prey density. A coexistence equilibrium must satisfy \label{defQbar} and the conditions
\be
f(X,y,\theta_i) = p_i y g(\tilde Q).
\label{PreyConditions}
\ee 
Let $\tilde X$ be defined as the solution of 
$$f(\tilde X,\tilde y,\theta_2) = \tilde y g(\tilde Q).$$
We have $f(\theta_2,0,\theta_2)=0$, so $\tilde X = \theta_2 + O(\tilde y)$. 
Then condition \eqref{PreyConditions} for prey type 1 can be
satisfied by choosing $\theta_1 = \theta_2 + O(\tilde y)$ such that 
$$f(\tilde X,\tilde y,\theta_1) = p_1 y g(\tilde Q)$$.

The last thing needing to be shown is that there exist $\tilde x_1 >0, \tilde x_2>0$ such that 
$$\tilde x_1 + \tilde x_2 = \tilde X, \qquad p_1 \tilde x_1 + \tilde x_2 = \tilde Q.$$ 
This is a system of two linear equations in the two unknowns $\tilde x_1, \tilde x_2$ whose
solution (by inverting a $2 \times 2$ matrix) is  
\be
\left[ {\begin{array}{*{20}c}
{\tilde x_1 }  \\
{\tilde x_2 }  \\
\end{array} } \right] = \frac{1}
{{1  - p_1 }}\left[ {\begin{array}{*{20}c}
{\tilde X - \tilde Q}  \\
{\tilde Q - p_1 \tilde X}  \\
\end{array} } \right].
\label{Generalxbars}
\ee
For $\tilde y$ small, $\tilde X \approx \theta_2 > \tilde Q$ so $\tilde x_1 >0$ in \eqref{Generalxbars}. 
And for $p_1$ sufficiently small, $p_1 \tilde X < \tilde Q$ so $\tilde x_2 >0$. So there do exist
positive prey steady-states giving a coexistence equlibrium for $\tilde y$ near 0, and their
limiting value (as $\tilde y \to 0$ and the corresponding $\theta_1 \to \theta_2$) is given
by \eqref{Generalxbars} with $\tilde X = \theta_2$. 

The argument above also identifies when $p_1$ is ``sufficiently small'' for the limiting 
values of $\tilde x_1, \tilde x_2$ to be positive as $\theta_1 \to \theta_2$. The limiting value of $\tilde x_1$ is
always positive, and the limiting value of $\tilde x_2$ is positive if $p_1 \tilde X - \tilde Q >0$
in the limit, i.e. if $p_1 \theta_2 < \tilde Q$. Thus, the coexistence
equilibrium region extends up to the line $\theta_1 = \theta_2$ so long as   
$  p_1 < p^* = \tilde Q/\theta_2.
%\label{Generalpstar}
$

\section{Discussion} 
\label{Discussion}
The model studied in this paper is three dimensional, with a few fairly tame nonlinearities -- just like the Lorenz equations.  So it is not surprising that a complete mathematical analysis of the model has not been possible. Nonetheless, we have come a long way towards our goal of characterizing how and when rapid evolution can affect the ecological dynamics resulting from predator-prey interactions. 

Our primary questions concern the generality of the phenomenon of ``evolutionary" limit cycles in  
predator-prey interactions, and the conditions in which such cycles might be observed. 
A combination of analysis and numerical studies suggests that evolutionary dynamics are not omnipresent, 
but neither are they knife-edge phenomena existing only in a narrow range of
parameter values. Instead, the types of cycles observed by Yoshida et al.  
\cite{Yoshida2003,Yoshida2006} are both robust and general. 
They occur in a specific but substantial and 
biologically relevant region of the parameter space, and in a general class of predator-two prey models
that includes a two-prey model with Lotka-Volterra prey competition terms  
\cite{AbramsMatsuda1997,Abrams1999,Kretzschmar1993}, and the standard two
prey chemostat model \cite{Butler1986,JonesEllner2004,Yoshida2003} with mechanistic modeling
of resource competition between the prey.

We have shown that evolutionary cycles arise through a bifurcation from a stable coexistence equilibrium,  
that occurs when defense against predation for the defended type remains relatively inexpensive but nevertheless becomes very effective. 
Cryptic population dynamics, where the predator cycles
but the total prey density remains nearly constant, occur as a limiting case when effective defense  
comes at almost zero cost \cite{Yoshida2006}. These regions in parameter space are biologically relevant because empirical studies
have shown that defense - be it against predation or against antimicrobial compounds - can arise 
quickly and can be both highly effective and very cheap 
\cite{Andersson1999,Gagneux2006,Yoshida2004}.  
For example, Gagneux et al. 
\cite{Gagneux2006} showed that in laboratory cultures of 
{\it Mycobacterium tuberculosis} (TB) mutants, prolonged treatment
with antibiotics results in multi-drug resistant strains of TB
with no fitness costs for resistance, and furthermore that most
naturally circulating resistant TB strains are either low or no
cost types.  Indeed, fitness tradeoffs in the production of defensive
structures and compounds are notoriously difficult to demonstrate, and in many
empirical studies, no fitness tradeoff was actually found 
\cite{Andersson1999,Bergelson1996,Strauss2002}.

We close by listing some open questions. ``Proving things is hard'' (Hal Smith,
\textit{personal communication}),  but others may succeed where we have not. Concerning the
model in this paper, 
\begin{itemize} 
\item When does the Jacobian at a coexistence equilibrium have a pair of complex conjugate eigenvalues?  There will be 3 real, negative eigenvalues if the two prey types are very similar and the
interior equilibrium for the {(predator + vulnerable prey)} exists and is a stable node. 
However, our numerical results suggest the full system (at a coexistence
equilibrium) has complex conjugate eigenvalues 
whenever the {(predator+vulnerable prey)} system has an interior equilibrium with complex conjugate eigenvalues.
\item Can there be coexistence of the predator and both prey on a limit cycle or other attractor,
even when there is no coexistence equilibrium? Numerical evidence suggests that the answer is ``no'' for
the chemostat model: for $k_1$ below (above) the range of values at which a 
coexistence equilibrium exists, the defended (vulnerable) prey type outcompetes the other. 
As it is difficult to distinguish between persistence and slow competitive exclusion numerically, it is likewise hard to map reliably the parameter region where both prey coexist on a nonpoint attractor. 

\item  On the bifurcation curve 
$0 \leq p_1 \leq p^*, \quad k_1 =k_2$, the Jacobian of the general model
\eqref{pmc} has zero as a double root with algebraic multiplicity 2 and 
geometric multiplicity 1. 
Generically, this situation gives a {\em Takens-Bogdanov} bifurcation 
\cite{Kuznetsov1994}. Do the higher order conditions for Takens-Bogdanov,
which hold generically, hold for our model \eqref{eq1}?
   
\item A general two-prey, one-predator chemostat can exhibit a wider
range of dynamic behaviors than we have observed in a system where the prey differ only in their $p$ and $k$ values (see \cite{VayenisPavlou99} and references therein). Indeed, these predicted dynamics have been observed in other experimental systems \cite{Becks2005}.  The absence of some dynamics from our system, if true and verifiable,  could indicate a qualitative difference between within-species evolutionary dynamics resulting from prey genetic diversity, and food-web dynamics with one predator feeding on a several prey species whose within-species heritable variation is much smaller than the functional differences among prey species. 
\end{itemize} 
Finally, how robust are the phenomena of evolutionary and cryptic predator-prey cycles in more complex food webs involving multiple 
predator and prey species?

\newpage
\appendix
\section*{Appendices}
Appendices \ref{EigenPhase} and \ref{HopfConditions} summarize 
some general results useful to us here, and contain nothing original. 
In Appendix \ref{SSAppendix} we derive
the expressions for coexistence steady states in the reduced and rescaled two-prey
chemostat model, and in Appendix \ref{JacobianAppendix} we derive the Jacobian matrix
and prove that it has negative determinant at any coexistence steady state. In Appendix 
\ref{InvadeEdibleCycle} we derive the conditions in which a limit cycle of the 
{(predator + edible prey)} subsystem can can be invaded by the 
defended prey. Finally, in Appendix \ref{ECeigenvals}
we show generally that for realized cost $\theta_1$ sufficiently close to $\theta_2$ and 
$0 \le p_1 \le p^*$, the coexistence equilibrium for the general model \eqref{pmc} always has a pair of 
complex conjugate eigenvalues.

\section{Appendix: Eigenvectors and phase relations} \label{EigenPhase}
The contents of this Appendix appear to be well-known, but we have not 
seen them summarized anywhere in print. We consider oscillations in a linear system 
\be
\dot x = \mathbf{J}x
\label{linsys}
\ee    
resulting from the real matrix $\mathbf{J}$ having complex conjugate eigenvalues 
$$\lambda, \bar \lambda = a \pm ib \mbox{ with } b>0,$$ 
where $i=\sqrt{-1}$ and the overbar denotes complex conjugation. The corresponding eigenvectors are also a
complex conjugate pair $w, \bar w$. 
%because $\mathbf{J}w = \lambda w$ implies that
%$\mathbf{J}\bar w = \bar{(\mathbf{J}w)} = \bar{(\lambda w)} = \bar \lambda \bar w.$
The resulting oscillatory terms in solutions of \eqref{linsys} are of the general form
$A e^{\lambda t}w + B e^{\bar \lambda t}\bar w.$
In order for these to be real (as solutions of \eqref{linsys} must be), we must 
have $B= \bar A$. Then writing $A=r e^{i \theta}, r>0$, the solutions are
proportional to 
\be
z(t) \equiv e^{i \theta} e^{ibt}w + e^{-i \theta}e^{-ibt}\bar w. 
\label{oscterms2}
\ee
We are interested in the relative phases of the oscillations by different components
in $z(t)$. Write 
$
w_j = r_j e^{i \phi_j}
$
for the $j^{th}$ component of $w$. The $j^{th}$ component of $z(t)$ is then 
\be 
r_j (e^{i(\phi_j + \theta + bt )} + e^{-i(\phi_j + \theta + bt)}) = 2 r_j \cos (\phi_j+ \theta+ bt).
\ee  
The relative phases of the $j^{th}$ and $k^{th}$ components in solutions proportional to $z(t)$ 
is therefore given by $\phi_j - \phi_k.$ When this is near 0 components $j$ and $k$ are 
oscillating in phase, and when it is near $\pm \pi$ they are oscillating nearly out of phase. 

We are interested in the phase difference between the predators and total prey density. For 
that we can use a linear change of variables 
\[
\left[ {\begin{array}{*{20}c}
   u  \\
   v  \\
   y  \\
 \end{array} } \right] = \left[ {\begin{array}{*{20}c}
   {x_1  + x_2 }  \\
   {x_1  - x_2 }  \\
   y  \\
 \end{array} } \right] = \mathbf{A}\left[ {\begin{array}{*{20}c}
   {x_1 }  \\
   {x_2 }  \\
   y  \\
 \end{array} } \right], \quad \mathbf{A} = \left[ {\begin{array}{*{20}c}
   1 & 1 & 0  \\
   1 & { - 1} & 0  \\
   0 & 0 & 1  \\
 \end{array} } \right].
\]
In transformed coordinates the Jacobian matrix becomes $\mathbf{AJA}^{-1}$, and Jacobian eigenvectors $w$
are transformed to $\mathbf{A}w$. The dominant eigenvector component for $x_1+x_2$ is therefore the sum  
of the components for $x_1$ and $x_2$.  

\section{Appendix: Stability conditions} \label{HopfConditions}
In this Appendix we review criteria for local stability of equilibria in a three-dimensional 
system of ordinary differential equations.

The diagonal expansion (\cite{Searle1982}, section 4.6) is an expression for $\det(A+D)$ where $A$ is square
and $D$ is diagonal. For $D=xI$ and $A$ of order $n$ it states that
\be
\det(A+xI)=x^n + x^{n-1}T_1(A)+x^{n-2}T_2(A)+ \cdots + T_n(A)
\label{DiagExpand}
\ee
where $T_j(A)$ is the sum of all principal minors of order $j$ (a principal minor of order $j$
is the determinant of a $j \times j$ submatrix of $A$ whose diagonal is a subset of the diagonal of $A$ -- that
is, a submatrix obtained by selecting $n-j$ diagonal elements of $A$ and deleting the row and column
containing that element). Note that $T_n(A)=\det(A)$ and $T_1(A)=\mbox{trace}(A)$. 

For a $3 \times 3$ matrix the characteristic polynomial is 
\be
p(\lambda) \equiv \det(\lambda I-A) = \lambda^3 + c_2 \lambda^2 + c_1 \lambda + c_0.  
\label{CharPoly3}
\ee
Comparing with \eqref{DiagExpand} and noting that and that $Tr_j(-A)=(-1)^j Tr_j(A)$,
we have 
\be
c_0=T_3(-A)=-\det(A),\quad c_1 = T_2(-A)=T_2(A), \quad c_2=\mbox{trace}(-A)=-\mbox{trace}(A).
\ee 
In the notation of \eqref{CharPoly3}, the Routh-Hurwitz stability criteria for 
order-3 systems (May 1974) is 
\be
 c_0 > 0, \quad c_1>0, \quad c_2 > 0, \quad c_1 c_2 > c_0. 
\label{RH3}
\ee
Loss of stability through a Hopf bifurcation occurs when the third condition in \eqref{RH3} is
violated, with the $c_i$ all positive \cite{Guckenheimer1997}. 

\section{Coexistence steady states for the rescaled chemostat model} 
\label{SSAppendix}
We consider here the two-prey model \eqref{eqs6} 
Setting $\dot y=0$ and solving gives the steady state value of $Q$, 
$\tilde Q = \frac{k_b}{g-1}$. We solve for $\tilde X$ and $\tilde y$ as follows.  
Defining $Z=1-X-y$ and noting that $\frac{g}{k_b + \tilde{Q}} = \frac{1}{\tilde{Q}},$ the
conditions $\dot x_1=\dot x_2=0$ imply 
\be
\frac{m\tilde{Z}}{k_1 + \tilde{Z}}- \frac{p_1\tilde{y}}{\tilde{Q}} = 
\frac{m\tilde{Z}}{k_2 + \tilde{Z}}- \frac{\tilde{y}}{\tilde{Q}} =1. 
\label{App2}
\ee
Solving \eqref{App2} for $\tilde{y}$ gives two expressions which 
remain equal within the coexistence region:
\be
\tilde{y} = \frac{\tilde{Q}}{p_1} \left[ \frac{(m -1)\tilde{Z} - k_1 }{k_1 + \tilde{Z}} \right ], \quad 
\tilde{y} =  \tilde{Q}\left[\frac{(m -1)\tilde{Z} - k_2 }{k_2 + \tilde{Z}} \right].
\label{App3}
\ee
Setting the two expressions for $\tilde{y}$ equal, we can solve for $\tilde{Z}$:
\be
\label{App4}
\tilde{Z} = \frac{1}{2(1-p_1)(m-1)}\left[\zeta + \sqrt{\zeta^2 + 4(m-1)(1-p_1)^2k_1k_2}\right]
\ee
where 
$$
\zeta=k_1\left(1 + p_1(m-1)\right) - k_2\left((m-1) + p_1\right).
$$
Finally, recalling that $\tilde{Z}=1-\tilde{X}-\tilde{y}$, then $\tilde X= 1-\tilde Z - \tilde y.$
Expressions for $\tilde{x_1}$  and $\tilde{x_2}$ in terms of $\tilde X$ and $\tilde Q$
are derived and shown in the text. 

\section{Jacobian at a coexistence equilibrium} \label{JacobianAppendix}
The general expression \eqref{GeneralJacobian} for Jacobian entries at
a coexistence equilibrium implies that all entries in the $i^{th}$ row
of the Jacobian have common factor $\tilde x_i$,
so $\det(J)=\tilde x_1 \tilde x_2 \tilde y \det(\tilde J)$ where 
$\tilde J(i,j)= \frac{\partial \tilde r_i}{\partial x_j}$ with $x_3=y$. 
Let $\tilde F$ denote the steady state per-capita feeding rate for the predator, 
\be
\tilde F=\frac{1}{k_b+p_1 \tilde x_1 + p_2 \tilde x_2},
\ee 
and the $a_i$ are defined by \eqref{defnAi} with $\tilde Z = 1 - \tilde x_1 - \tilde x_2 - \tilde y$; 
equation \eqref{App4} gives the general expression for $\tilde Z$.   

Taking the necessary partial derivatives,   
\be
\tilde J  = \left[ {\begin{array}{*{20}c}
   {- a_1 + g p_1^2 \tilde y \tilde F^2 } & {- a_1 + g p_1 p_2 \tilde y \tilde F^2 } & {-a_1 - gp_1\tilde F}  \\
   {- a_2 + g p_1 p_2 \tilde y \tilde F^2 } & { -a_2 + g p_2^2 \tilde y \tilde F^2 } & {-a_2 - gp_2\tilde F}  \\
   {p_1 g k_b \tilde F^2} & {p_2 g k_b \tilde F^2} & 0  \\
\end{array} } \right].
\ee

We now show that the determinant of the Jacobian is always negative for the
general model \eqref{pmc}, and therefore for the chemostat model, unless $p_1=p_2$. 
For \eqref{pmc} with the scaling $p_2=1$ we have 
%
%The general expression \eqref{GeneralJacobian} for Jacobian entries at
%a coexistence equilibrium, which applies also to the general model \eqref{pmc}, 
%implies that all entries in the $i^{th}$ row of the Jacobian have common factor
%$\tilde x_i$,
%so $\det(J)=\tilde x_1 \tilde x_2 \tilde y \det \tilde J$ where $\tilde J(i,j)= \bar
%r_{i,j}$. 
%Then taking the necessary partial derivatives with the scaling $p_2=1$, we get 
\be
\tilde J = 
\begin{bmatrix} 
f_X - p_1^2 \tilde y \tilde g' & f_X - p_1 \tilde y \tilde g' & f_y - p_1 \tilde g \\
f_X - p_1 \tilde y \tilde g' & f_X -  \tilde y \tilde g' & f_y -  \tilde g \\
\tilde h' p & \tilde h' & 0 \\
\end{bmatrix}
\ee
where $\tilde g = g(\tilde Q), \tilde g' = g'(\tilde Q)$ and 
$\tilde h' = h'(\tilde Q), h(Q)=Qg(Q)$. 
Then using basic products of determinants, $\det (\tilde J)$ equals 
\begin{equation} 
\tilde{h'}  \left| {\begin{array}{*{20}c}
   {f_X } & {f_X } & {f_y  - p_1 \tilde g}  \\
   {f_X } & {f_X } & {f_y  - p_1 \tilde g}  \\
   {p_1 } & 1 & 0  \\
\end{array} } \right| 
= \tilde{h'} \left| {\begin{array}{*{20}c}
   {f_X } & {f_X } & {f_y  - p_1 \tilde g}  \\
   0 & 0 & {(p_1  - 1)\tilde g}  \\
   {p_1 } & 1 & 0  \\
\end{array} } \right| = (1-p_1)^2 \tilde{h'} \tilde{g} f_X
\end{equation}
which is negative (unless $p_1=1$) because $\tilde h'>0, \tilde g>0$ and $f_X<0$.

\section{Appendix: Invasion of an edible prey limit cycle} 
\label{InvadeEdibleCycle}
Following \cite{Abrams1999} we give here the condition for invasion of a {predator + edible prey}
limit cycle by a rare defended prey type. Along the limit cycle we have 
$\bracket{\dot \log y} =0$ and therefore  $\bracket{\frac{g x_2}{k_b+x_2}}=1.$ By Jensen's
inequality, this implies that $\frac{g \bracket{x_2}}{k_b + \bracket{x_2}} > 1$, and 
therefore $\bracket{x_2} > \tilde Q$. We also have $\bracket{\dot {\log x_2}} =0$ 
along the limit cycle, so 
\begin{equation}
\bracket{\frac{gy}{k_b+x_2}} =1 + \bracket{\frac{m(1-x_2-y)}{k_2 + 1 - x_2 - y}}.
\label{x2dot}
\end{equation}
A rare defended prey can invade if $\bracket{\dot {\log x_1}}>0$, i.e. if 
$$ 0< \bracket{\frac{m(1-x_2-y)}{k_1 + 1- x_2 -y} - p_1\frac{gy}{k_b+x_2}- 1}= \bracket{\frac{m(1-x_2-y)}{k_1 + 1- x_2 -y}} - p_1 \bracket{\frac{gy}{k_b+x_2}}- 1$$
Using \eqref{x2dot} and simplifying, we get the invasion condition
in terms of $p_1, k_1$):
\be
\label{Invadelimitcycle}
p_1  < \frac{\bracket{ \zeta(k_1) } -1}{\bracket{ \zeta(k_2) } + 1},
\ee
where $$
\zeta(k_i) = \frac{m(1-x_2-y)}{k_i + 1-x_2 - y}.
$$
Note that the right-hand side of \eqref{Invadelimitcycle} can be 
computed for all $k_1$ using one long simulation of the (predator + vulnerable prey) system, and yields $p_1$ as a function of $k_1$.

\section{Appendix: Eigenvalues for $\theta_1 \uparrow \theta_2, p_1 \le p^*$} 
\label{ECeigenvals}

We show here that for $\theta_1$ sufficiently close to $\theta_2$ and  $0 \le p_1
\le p^*$
in the general model \eqref{pmc}, the coexistence equilibrium always has a pair of
complex
conjugate eigenvalues. As $\theta_1 \to \theta_2$, in this range of $p_1$ values
$\tilde y \to 0$, so we
set $\tilde y=\epsilon \ll 1$ and use a series expansion in $\epsilon$ 
of the characteristic polynomial (i.e. we regard $\theta_1$ as a function of $\tilde y$ with
all else held fixed, rather than \textit{vice versa}). The Jacobian at the
coexistence equilibrium is an
$O(\epsilon)$ perturbation of \eqref{LimitJacobian} and so to leading order has the
form 
\begin{equation}
J(\epsilon)=
\begin{bmatrix}
A + \epsilon a_{11} & A + \epsilon a_{12} & B + \epsilon a_{13} \\
C + \epsilon a_{21} & C + \epsilon a_{22} & D + \epsilon a_{23} \\
\epsilon a_{31} & \epsilon a_{32} & 0 \\
\end{bmatrix}
\label{Jepsilon}
\end{equation}
with $A,B,C,D <0$, and $a_{31}=p_1 a_{32} > 0$ (the last holds because
$\dot y/y$ is a function of $Q=p_1 x_1 + x_2$ with the scaling $p_2=1$). 
$J(0)$ has eigenvalues zero (with algebraic multiplicity 2) 
and $A+C<0$, and we need to approximate the near-zero eigenvalues for $\epsilon$ small.
The characteristic polynomial of $J(\epsilon)$ is a cubic in $\lambda$ but the 
near-zero eigenvalues are at most $O(\sqrt{\epsilon})$, so for our purpose the
$\lambda^3$ terms in the 
characteristic polynomial can be neglected. This leaves a quadratic
polynomial in $\lambda$, which will have complex conjugate roots if its discriminant
is negative. Using Maple to compute the characteristic polynomial of \eqref{Jepsilon},
discard $\lambda^3$ terms and expand the remainder about $\epsilon=0$, to 
leading order in $\epsilon$ the discriminant is 
$$ 4\epsilon (a_{32}-a_{31})(A+C)(AD-BC) $$
which will be negative if $AD-BC>0$. Referring to \eqref{LimitJacobian} some algebra
gives 
$$AD-BC = \tilde x_1 \tilde{x}_2 \tilde{f}_X \tilde g (p_1 -1)$$
which is positive because $f_X <0$, as desired.

\end{document}